\documentclass[12pt]{article}
\usepackage[utf8]{inputenc}
\usepackage[margin=0.8in]{geometry}
\usepackage{blindtext, xfrac}
\usepackage[english]{babel}
\usepackage[T1]{fontenc}
\usepackage{titling}
\usepackage{amsmath}
\usepackage{comment}
\usepackage{amssymb}
\usepackage{cite}
\usepackage{graphicx}
\usepackage{booktabs}
\usepackage{physics}
\usepackage{nameref}
\usepackage{ bbold }
\usepackage{textgreek}
\usepackage{hyperref}
\usepackage{gensymb}
\usepackage{float}
\usepackage[usenames,dvipsnames]{color}
\usepackage{svg}
\usepackage{placeins}
\usepackage{subcaption}

\definecolor{greeny}{rgb}{0.2,0.8,0.5}

\title{Secluded Dark Composites and Remnant Binding Fields}

\author{Katarina Bleau$^{1,2}$, Yilda Boukhtouchen$^{1}$, Joseph Bramante$^{1,3}$, and Rohan Kulkarni$^{1}$
\\
$^1$The Arthur B. McDonald Canadian Astroparticle Physics Research Institute, \protect\\ Department of Physics, Engineering Physics, and Astronomy, \protect\\ Queen's University, Kingston, Ontario, K7L 2S8, Canada
\\
$^2$PRISMA Cluster of Excellence \& Mainz Institute for Theoretical Physics, \protect\\ Johannes Gutenberg University, 55099 Mainz, Germany
\\
$^3$Perimeter Institute for Theoretical Physics, Waterloo, ON N2J 2W9, Canada}

\date{\today}

\begin{document}
\maketitle

\begin{abstract}

Dark matter may freeze-out and undergo composite assembly while decoupled from the Standard Model. In this secluded composite scenario, while individual dark matter particles may be too weakly-coupled to detect, the assembled composite can potentially be detected since its effective coupling scales with number of constituents. We examine models and observables for secluded composites, and in particular we investigate the cosmological abundance of the composite binding field, which is generated during freeze-out annihilation and secluded composite assembly. This binding field could be discovered as a new relativistic species in the early universe or through later interactions as a subdominant dark component.

\end{abstract}

\section{Introduction}

Dark matter's identity and interactions remain unsolved mysteries in the modern scientific era. Amongst the many possible dark matter candidates, composite dark matter has some unique formation pathways and interactions, which are still being explored\cite{Witten:1984rs,Farhi:1984qu,DeRujula:1984axn,Goodman:1984dc,Drukier:1986tm,Nussinov:1985xr,Bagnasco:1993st,Alves:2009nf,Kribs:2009fy,Lee:2013bua,Laha:2013gva,Laha:2015yoa,Krnjaic:2014xza,Detmold:2014qqa,Jacobs:2014yca,Wise:2014ola,Hardy:2014mqa,Hardy:2015boa,Bramante:2018qbc,Gresham:2018anj,Bramante:2018tos,Ibe:2018juk,Grabowska:2018lnd,Coskuner:2018are,Bai:2018dxf,Digman:2019wdm,Bai:2019ogh,Bramante:2019yss,Bhoonah:2020dzs,Clark:2020mna,Gresham:2017zqi,Gresham:2017cvl,Wise:2014jva,Acevedo:2020avd,Acevedo:2021kly,Fedderke:2024hfy,Dhakal:2022rwn,Moore:2024mot,Xie:2024mxr}.
For example, it was recently shown that a light scalar field, which produces an attractive potential that binds together fermionic composites, can produce multiple low-energy recoils in underground dark matter search experiments \cite{Acevedo:2021kly} and even Bremsstrahlung radiation from Standard Model nuclei accelerated in the dark matter composite's interior \cite{Acevedo:2020avd}. This and other novel searches for composite dark matter (e.g.~\cite{Jacobs:2014yca,Bramante:2018qbc,Gresham:2018anj,Bramante:2018tos,Ibe:2018juk,Grabowska:2018lnd,Bai:2020jfm,Croon:2020wpr,Das:2021drz,Acevedo:2024lyr,Moore:2024mot,Bramante:2024hbr}) motivate a deeper understanding of its origins.

In this work, we explore the dynamics of secluded composite dark matter in the early universe, focusing on composite dark matter made of a dark matter Dirac fermion $\chi$ and a light scalar field $\varphi$, which we will also refer to as a “remnant binding field.” The formation of composite dark matter particles in this framework involves assembly of fermionic composites with associated emission of the light scalar field, leading to a leftover population of  scalar fields in the late universe, which may be discovered through cosmological observations. Secluded dark matter, where dark matter's abundance is determined while out of thermal and chemical equilibrium from the Standard Model, has been considered previously \cite{Pospelov:2007mp,Feng:2008ya,Feng:2008mu,Evans:2017kti}, but as we will explore in this work, there are some unique properties and signatures associated with secluded composite dark matter.

The rest of this paper is organized as follows. In Section \ref{sec:2}, we study the freeze-out dynamics of secluded composite dark matter, focusing on the annihilation process $\chi + \bar{\chi} \to \varphi + \varphi$, the subsequent formation of bound states, and the energy density evolution of the remnant scalar field $\varphi$ during composite assembly. We examine the contributions of $\varphi$ to cosmological observables including the effective number of relativistic species ($N_{\mathrm{eff}}$), and detail constraints from Planck data and projected CMB-S4 sensitivities. In Section \ref{sec:3}, we introduce a UV thermal coupling which ensures DM-SM thermal equilibrium during the early universe and sets the DM cosmological abundance. Further, we introduce a Higgs portal interaction to connect the dark sector to the Standard Model. This interaction allows us to derive bounds on the mass and coupling of the Higgs portal scalar. We then use these results to place constraints on direct detection prospects for this model, considering both astrophysical and experimental limits. Section \ref{sec:4} concludes with a summary of our results and a discussion of future directions. Throughout this work, we use natural units where $\hbar = c = k_B = 1$.

\section{Secluded composite dark matter} \label{sec:2}
In this study we consider dark matter composed predominantly of fermions $\chi$, which assemble into composites under the influence of an attractive potential provided by a light scalar field $\varphi$. The freeze-out and composite assembly of dark matter both occur out-of-equilibrium with Standard Model radiation in the early universe. The Lagrangian for dark matter in this cosmology is 
\begin{equation}
    \mathcal{L}_{\mathrm{DM}} = \bar{\chi}(i\gamma^{\mu}\partial_{\mu}-m_{\chi})\chi + \frac{1}{2}\partial_{\mu}\varphi\partial^{\mu}\varphi - \frac{1}{2}m_{\varphi}^2\varphi^2 + g_{\chi\varphi}\bar{\chi}\varphi\chi + \mathcal{L}_{\mathrm{DM-SM}}, \label{eq:lagrangian_ds}
\end{equation}
where $\mathcal{L}_{\mathrm{DM-SM}}$ indicates additional terms coupling the dark matter fields to the Standard Model (SM). Section \ref{sec:3} will detail a feeble, low energy SM coupling to the dark sector via a Higgs portal mixing term. In addition, Section \ref{sec:3} will describe the coupling of both SM and DM fields to a heavy scalar, which equilibrates dark matter and Standard Model radiation at high temperatures, before they fall out of thermal and chemical equilibrium at lower temperatures, at which point the dark sector fields are referred to as ``secluded.''

In this Section we will assume the dark matter is out-of-equilibrium with the Standard Model, and in this state determine cosmological freeze-out and composite assembly dynamics for the dark matter fields given in Equation \eqref{eq:lagrangian_ds}. We will be particularly interested in the abundance of $\varphi$, which can be a light field that contributes to the number of relativistic degrees of freedom, or to the amount of matter-like energy in the early universe, depending on its mass. We will want  to determine how the energy density of $\varphi$ evolves in the early Universe. The abundance of $\varphi$ can depend both on production during freeze-out annihilation processes ($e.g.$ $\chi + \bar{\chi} \rightarrow \varphi + \varphi$), and the formation of two-body and N-body composites, which decay into their ground states by emitting $\varphi$. The abundance of $\varphi$ after the formation of secluded composites is one of the main focuses of this study.

\subsection{Secluded asymmetric dark sector abundance}
\label{sec:asym}

We now turn to the early Universe abundance of asymmetric dark matter fermions $\chi$ and the scalar field $\varphi$, which will later on be responsible for binding together our composite dark matter states. We will specifically be interested in \textit{asymmetric} dark matter fermions, since asymmetric composites will be longer lived than a composite state formed of both particles and antiparticles (which would tend to annihilate). We assume that the asymmetric $\chi/ \bar \chi$ abundance is set by a high temperature asymmetry generation mechanism, see e.g.~\cite{Petraki:2013wwa,Zurek:2013wia} for a review of asymmetric dark matter and \cite{Affleck:1984fy,Bramante:2017obj} for discussion of high mass asymmetry generation mechanisms. 

In the case that dark sector fermions are secluded from SM radiation, freeze-out depletion of $\chi \bar \chi$  will occur predominantly through annihilation to the light scalar field via the process $\chi + \bar{\chi} \rightarrow \varphi + \varphi$, leaving a residual $\chi$ population determined by the initial $\chi / \bar{\chi}$ asymmetry. For the composite dark matter models we will consider, there will also be a depletion of dark matter's effective mass associated with composite formation, since the dark matter field's mass outside composites, $m_\chi$ will be larger than $\bar m_\chi$, the mass inside dark matter composite states. 

We will evolve the number density of dark matter ($n_{\chi}$), of the Standard Model ($n_{\mathrm{SM}}$) and of the light scalar ($n_{\varphi}$) as a function of the scale factor $a$ using standard Boltzmann equations. For convenience, we split the Boltzmann equations for $\chi$ into a symmetric component $n_{\chi}$ and an asymmetric residual component $n_{\chi,\mathrm{res}}$, which will be left after the symmetric component annihilates away during freeze-out. The Boltzmann equations are then given by
\begin{equation}
    aHn_{\chi}' + 3Hn_{\chi} = -n_{\chi}^2 \langle\sigma v\rangle + g_{\chi}^2 \left(\frac{m_{\chi} T}{2\pi}\right)^{3} e^{-2m_{\chi}/T} \langle\sigma v\rangle,
\label{eq:boltz_chi}
\end{equation}
\begin{equation}
    aHn_{\chi,\mathrm{res}}' + 3Hn_{\chi,\mathrm{res}} = 0,
\label{eq:boltz_res}
\end{equation}
\begin{equation}
    aHn_{\mathrm{SM}}' + 3Hn_{\mathrm{SM}} = 0,
\label{eq:boltz_SM}
\end{equation}
\begin{equation}
    aHn_{\varphi}' + 3Hn_{\varphi} = n_{\chi}^2 \langle\sigma v\rangle - g_{\chi}^2 \left(\frac{m_{\chi} T}{2\pi}\right)^{3} e^{-2m_{\chi}/T} \langle\sigma v\rangle,
\label{eq:boltz_phi}
\end{equation}
where the Hubble constant is
\begin{equation}
    H^2 = \frac{8\pi}{3M_{\mathrm{pl}}^2} \rho_{\mathrm{tot}}
\end{equation}
and where $\rho_{\mathrm{tot}}$ is the sum of all energy densities. For the cosmological epochs we are interested in, the energy density will be predominantly that of the SM, $\rho_{\mathrm{tot}} \simeq \rho_{\mathrm{SM}}$, where the energy density of Standard Model radiation is given by $\rho_{\mathrm{SM}} = \frac{\pi^2}{30} g_{*\mathrm{SM}} T^4$, and where $g_{*\mathrm{SM}}(T)$ is the number of relativistic degrees of freedom in the Standard Model at temperature $T$. 

The annihilation cross section for the interaction $\chi+ \bar \chi \rightarrow \varphi +\varphi$ is
\begin{equation}
    \langle\sigma v\rangle = \frac{3 \pi \alpha^2}{8 m_{\chi}^2},
\end{equation}
in terms of the coupling constant $\alpha \equiv g_{\chi \phi}^2/4\pi$. We note that, given the assumption that composites will form from an asymmetric abundance of $\chi$ particles, the coupling $\alpha$ must be large enough to deplete the symmetric component of dark matter during freeze-out processes, as we detail in Subsection \ref{sec:bound_state}. Since the comoving entropy density of the Universe remains conserved and the Universe is dominated by Standard Model radiation in this scenario, the temperature  $T$  can be expressed in terms of the scale factor  $a$  using the relation \mbox{$T(a) = g_{*\mathrm{SM}}(T(a=1))^{1/3}T(a=1)/g_{*\mathrm{SM}}(T(a))^{1/3}a$}. We assume the relativistic degrees of freedom of the Standard Model follow the standard evolution (see $e.g.$ the treatment of $g_{*\mathrm{SM}}(T)$ in \cite{Husdal:2016haj}), using this to numerically solve for $T$ at as a function of $a$. We assume that (prior to $\chi \bar \chi$ freezeout) the temperatures of  $\chi$  and  $\varphi$  are equal to that of the Standard Model, which is justified in the case that they started in equilibrium at high temperatures (a UV completion of this is detailed in Section \ref{sec:3}). Hence we define our initial conditions such that $\chi$ and $\varphi$ are decoupled from the Standard Model, but have the same initial conditions as one would expect from them being in thermal equilibrium with the Standard Model. In order to obtain matter-radiation equality (and the observed dark matter abundance) at the observed temperature, the initial number density of the asymmetric abundance of dark matter must be
\begin{equation}
    n_{\chi,\mathrm{res}}(a=1) = \frac{\pi^2g_{*\mathrm{SM}}T_{\mathrm{m-r}}T(a=1)^3}{30\bar{m}_\chi},
\end{equation}
where $T_{\mathrm{m-r}} = 0.8 \, \mathrm{eV}$ is the temperature at matter-radiation equality and $\bar{m}_{\chi} = m_{\chi} - BE$ is the dark matter mass bound in the composite, defined in terms of the binding energy $BE$ \cite{Acevedo:2020avd}. Henceforth, we will label the binding energy $BE_n(N)$ with two parameters: the internal excitation energy level $n$ and the number of constituents $N$.

\begin{figure}[htbp]
    \centering
      \begin{subfigure}[b]{0.45\textwidth}
        \includegraphics[width=\textwidth]{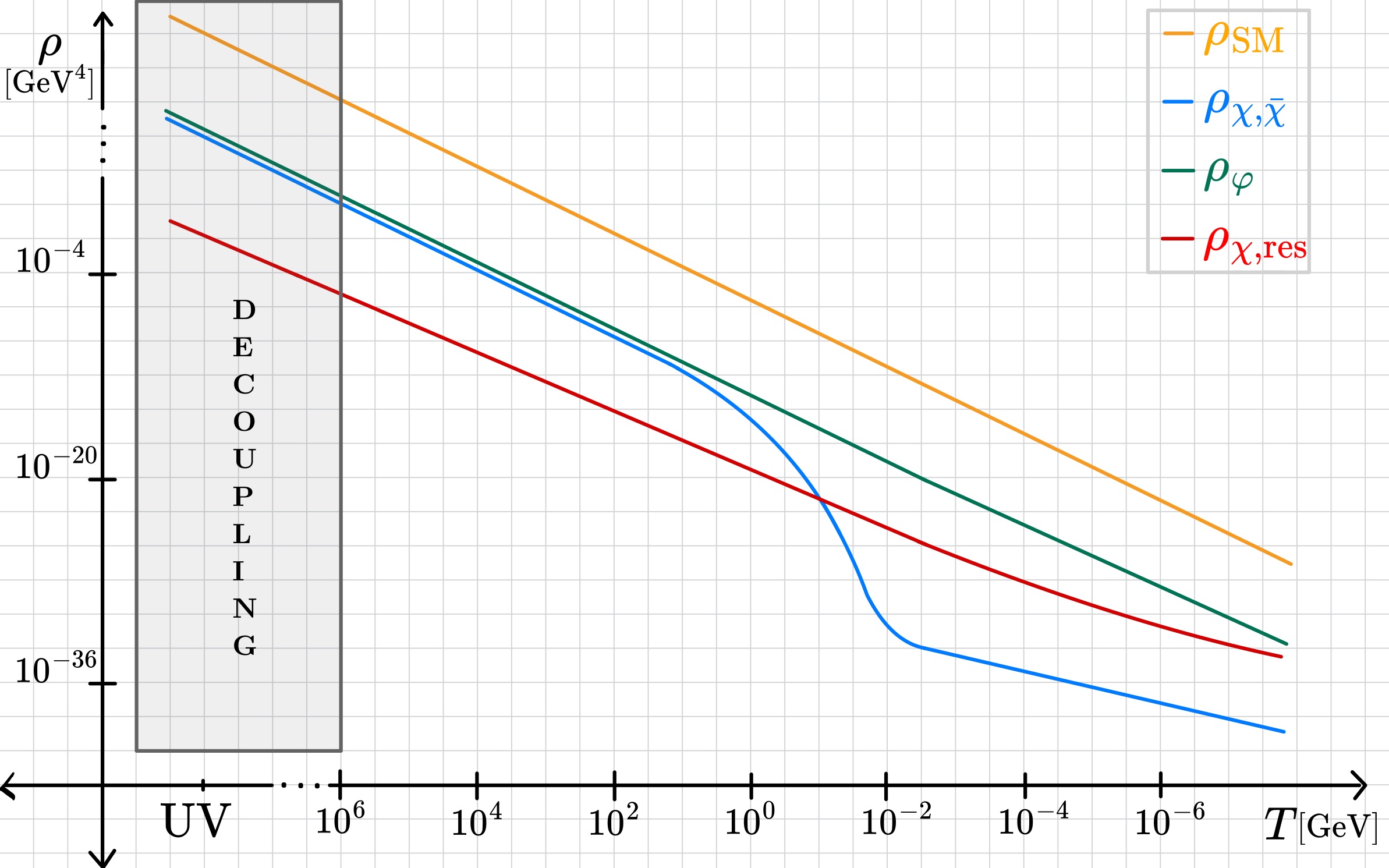}
        \caption{No bound state}
        \label{fig:sub1}
    \end{subfigure}
    \hfill
     \begin{subfigure}[b]{0.45\textwidth}
        \includegraphics[width=\textwidth]{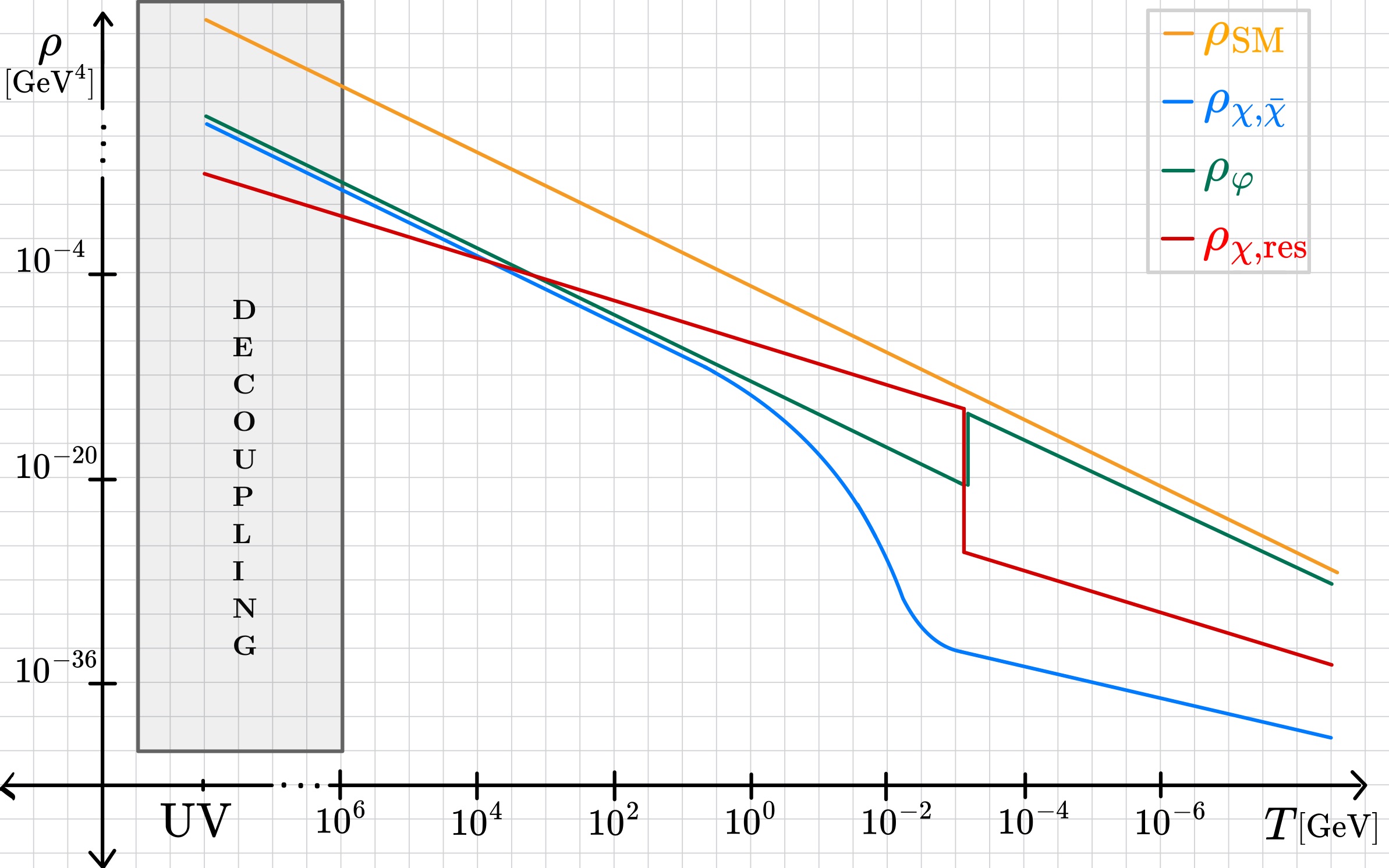}
        \caption{$m_\varphi = 10^{-3} \, \mathrm{eV}$}
        \label{fig:sub3}
    \end{subfigure}
    
    \vspace{\baselineskip} 
    
     \begin{subfigure}[b]{0.45\textwidth}
        \includegraphics[width=\textwidth]{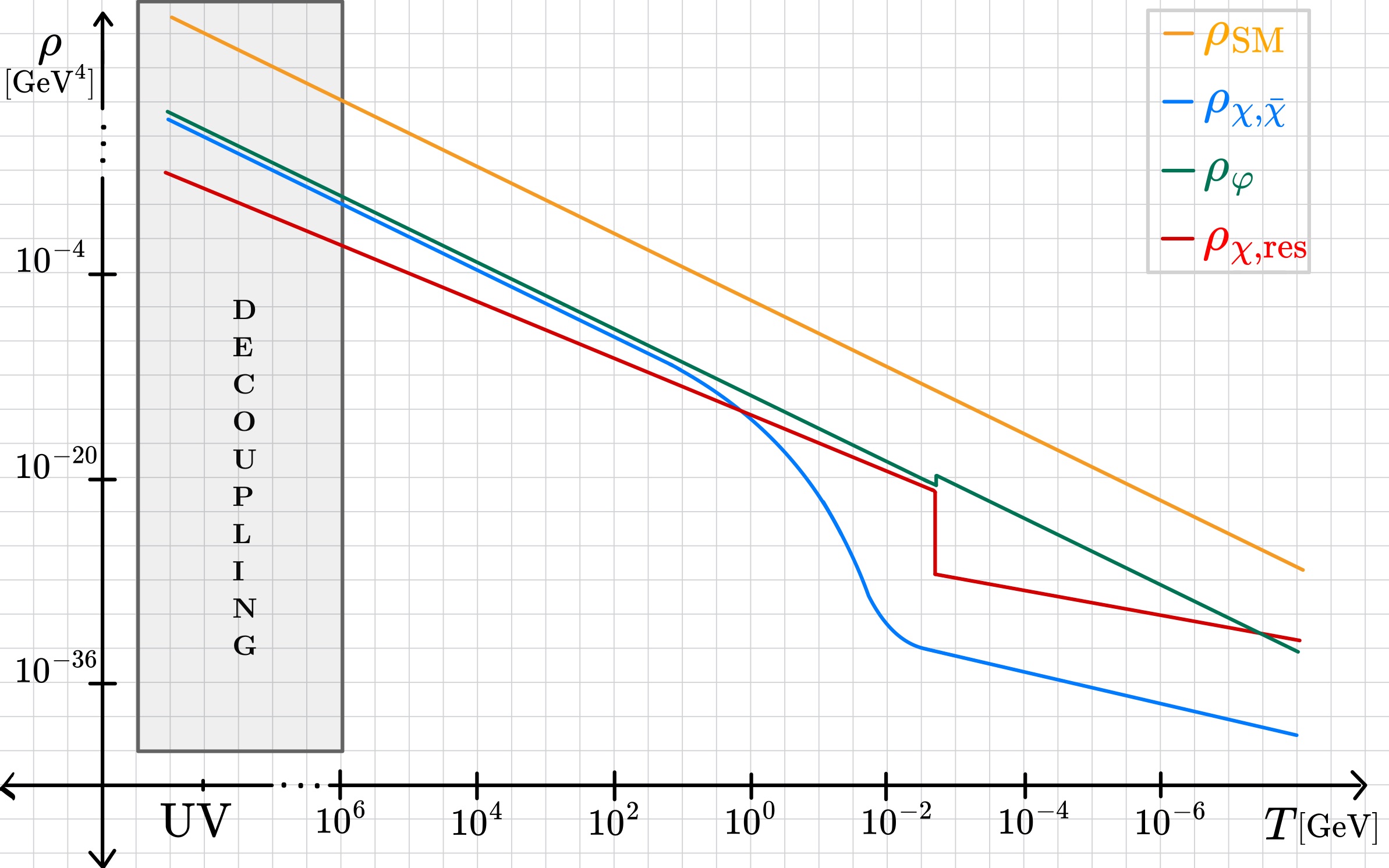}
        \caption{$m_\varphi = 1 \, \mathrm{eV}$}
        \label{fig:sub2}
    \end{subfigure}
    \hfill
       \begin{subfigure}[b]{0.45\textwidth}
        \includegraphics[width=\textwidth]{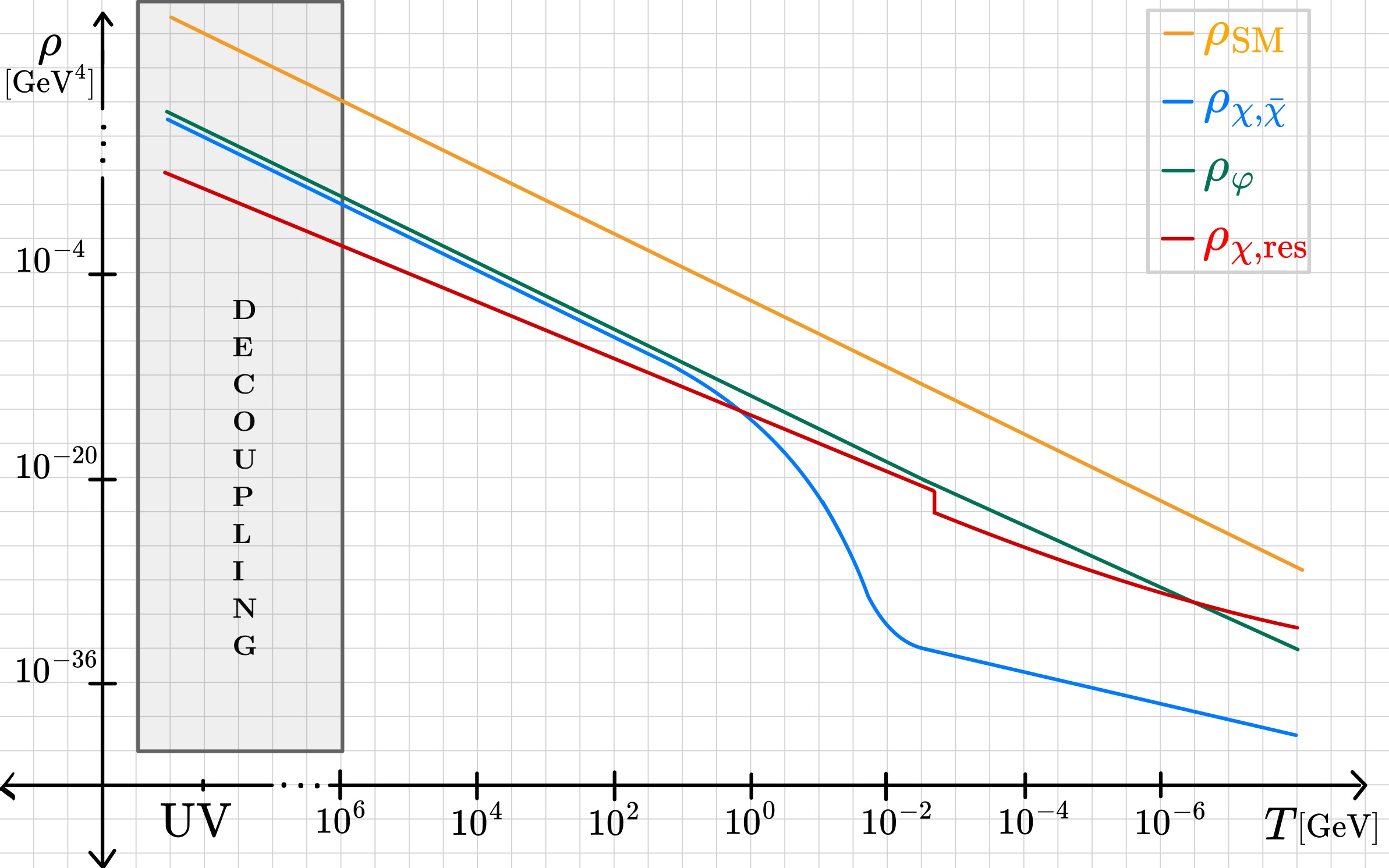}
       \caption{$m_{\varphi}=100 \, \mathrm{eV}$}
        \label{fig:sub4}
    \end{subfigure}
    \caption{Schematic cosmological energy densities, matching Boltzmann equation outputs, are shown as a function of the Standard Model radiation temperature $T$, for a secluded dark sector which can form composites out of  equilibrium with the Standard Model. The secluded dark matter field $\chi$ annihilates solely to scalar field $\varphi$, leaving behind a residual dark matter density ``$\chi,$ res''. The energy densities of the symmetric dark matter, asymmetric dark matter residual, Standard Model radiation and light scalar $\varphi$ components are represented by the blue, red, orange, and green curves respectively. In the top left panel, we show the case where no DM composite forms for comparison. In all other panels, we set $m_X$ = 10 GeV, $\alpha$ = 0.1, and vary the mass of the light scalar $m_{\varphi}$ as indicated. The asymmetric residual population of $\chi$s form composites at a lower temperature, during which process the binding energy of the composites is emitted as $\varphi$. The initial residual dark matter density ``$\chi$, res'' is set so that matter-radiation equality is obtained at $T=0.8$ eV. The initial $\varphi$ and $\chi$ abundance are set to reflect having previously been in thermal equilibrium with the SM at high temperature. }
    \label{fig:overall}
\end{figure}

When computing abundances, we initialize our Boltzmann equations using $T(a=1)$ = 10$T_{\mathrm{fo}}$, where $T_{\mathrm{fo}}$ is the freeze-out temperature, which we define using the standard freeze-out condition,
\begin{equation}
    H = g_{\chi} \left(\frac{m_{\chi} T_{\mathrm{fo}}}{2\pi}\right)^{3/2} e^{-m_{\chi}/T_{\mathrm{fo}}} \langle\sigma v\rangle.
\end{equation}
Then, to calculate energy densities in terms of the number densities found from the Boltzmann equations, we use 
\begin{equation}
    \rho_{\chi} = n_{\chi} m_{\chi}; \quad
    \rho_{\chi,\mathrm{res}} = n_{\chi,\mathrm{res}} m_{\chi}; \quad
    \rho_{\mathrm{SM}} = \frac{\pi^4 T n_{\mathrm{SM}}}{30 \zeta(3)}; \quad
    \rho_{\varphi} = \frac{\pi^4 T n_{\varphi}}{30 \zeta(3)}.
\end{equation}
In the upper left panel of Figure \eqref{fig:overall}, we evaluate the energy densities of $\chi$, $\varphi$ and the Standard Model as functions of the Standard Model radiation temperature, and show the cosmological abundances in the case that asymmetric $\chi$s do not form composites.

\subsection{Bound state formation}
\label{sec:bound_state}

Next, we consider the formation and decay of composite states formed from asymmetric fermions $\chi$. The minimum coupling $\alpha$ between $\chi$ and $\varphi$ which allows for the creation of two-body fermion bound states \cite{Wise:2014jva} is
\begin{equation}
    \alpha \gtrsim 0.1 \left(\frac{m_{\chi}}{100 \,\text{GeV}}\right)^{1/3}.
\label{eq:alpha}
\end{equation}
After the secluded dark sector freeze-out detailed in the previous section, the asymmetric population of residual fermions form composites in either an excited state $n > 1$ or in their ground state $n=1$. These composites begin forming once the formation process becomes efficient compared to the dissociation process. We assume that composites form when the Universe reaches a temperature of $T_{\text{form}} \simeq BE_1(2)/10$. Here we have chosen $T_{\text{form}}$ to be characteristic of a freeze-out temperature for a thermalized process - this estimate has been shown to provide relatively accurate results for composites with constituents whose binding energy is close to the $\chi$ bare mass \cite{Gresham:2017cvl} - and hence this estimate should be valid for the parameters we will consider, but in the future this could be revisited for models with smaller binding energies \cite{Acevedo:2024lyr}. 

\begin{figure}
    \centering
    \includegraphics[width=\textwidth]{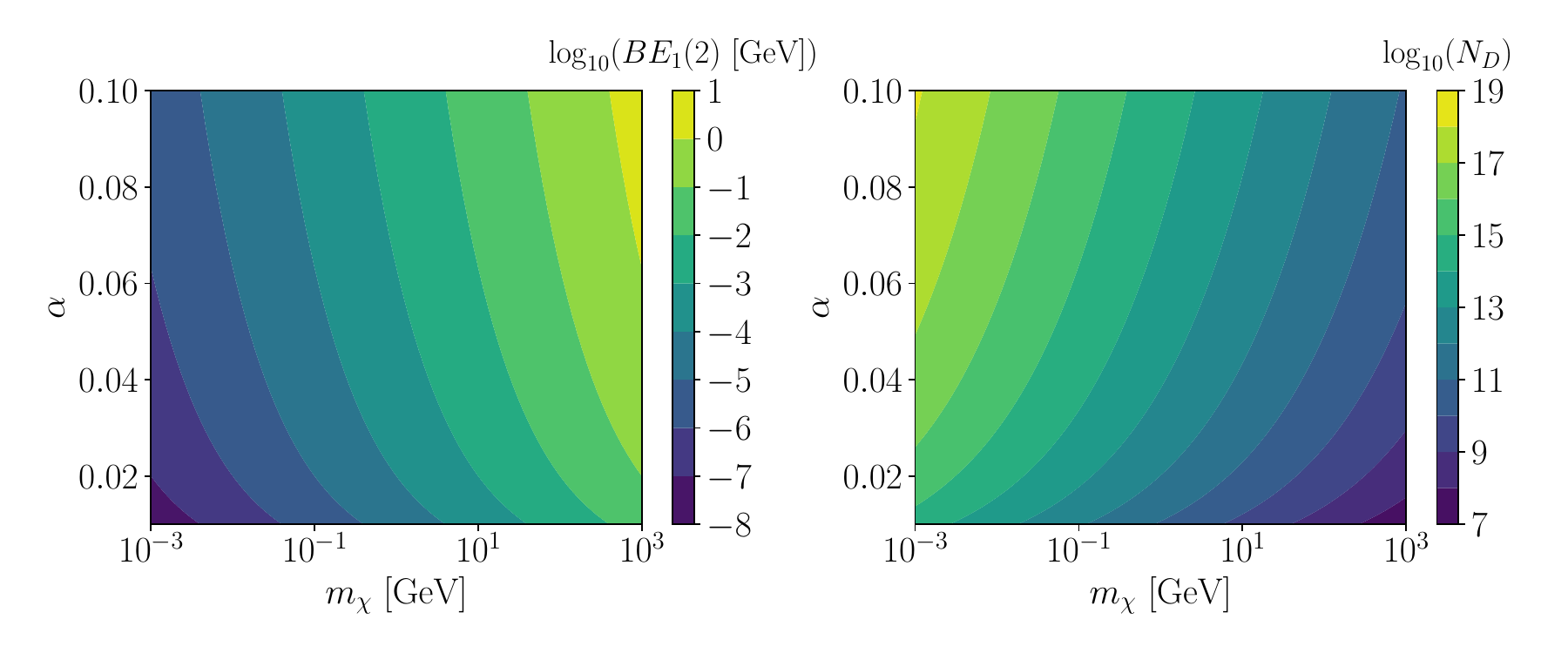}
    \caption{\textit{Left}: Binding energy (in log scale) of two-body bound state, from Equation \eqref{eq:BE_twobody}. \textit{Right}: Number of constituents (log scale) for a strongly-bound composite state, from Equation \eqref{eq:num-const}, assuming that the binding energy per constituent is $0.99 m_\chi$. (The actual binding energy per constituent, and the number of constituents per composite are given by Eqs. \eqref{eq:num-const}, \eqref{eq:eff_mass}.)}
    \label{fig:BE_and_NumConst}
\end{figure}

When the composites assemble, we will want to determine the time scale for them to decay to their ground states, $a_{\text{decay}}$. We will also want to examine how the formation of the two-body and $N$-body bound states and their decay into ground states affect the energy density of $\varphi$. In this scenario, the mass energy lost by the asymmetric fermions $\chi$ will be emitted as $\varphi$ particles. This energy results from the mass difference $m_\chi-\bar{m}_{\chi}$ of the fermion outside versus inside the composite. To account for this in a simple manner, we can define $\rho_{\chi,\mathrm{res}}$ and $\rho_{\varphi}$ as the following piecewise functions
\begin{equation}
\rho_{\chi,\mathrm{res}}(a) =
\begin{cases}
    n_{\chi,\mathrm{res}}(a) m_{\chi} & a < a_{\mathrm{form}} \\ n_{\chi,\mathrm{res}}(a) (m_{\chi}-BE_n(2)) & a_{\mathrm{form}} \leq a < a_{\mathrm{decay}} \\
    n_{\chi,\mathrm{res}}(a) (m_{\chi}-BE_1(2)) & a \geq a_{\mathrm{decay}},
\end{cases}
\label{eq:rho_chires_piecewise}
\end{equation}
\begin{equation}
\rho_{\varphi}(a) =
\begin{cases}
    \frac{\pi^4 T}{30 \zeta(3)}n_{\varphi}(a) & a < a_{\mathrm{form}} \\ \frac{\pi^4 T}{30 \zeta(3)}n_{\varphi}(a) + \frac{a_{\mathrm{form}}^4 n_{\chi,\mathrm{res}}(a_{\mathrm{form}}) BE_n(2)}{a^4} & a_{\mathrm{form}} \leq a < a_{\mathrm{decay}} \\
    \frac{\pi^4 T}{30 \zeta(3)}n_{\varphi}(a) + \frac{a_{\mathrm{decay}}^4 n_{\chi,\mathrm{res}}(a_{\mathrm{decay}}) BE_1(2)}{a^4} & a \geq a_{\mathrm{decay}}.
\end{cases}
\label{eq:rho_phi_piecewise}
\end{equation}
We note that in the above treatment, the comoving entropy density is not conserved. A more detailed treatment would involve introducing Saha relations into the Boltzmann equations. Nevertheless, in the case of two-body states, we can solve for the rate at which the composites will decay into their ground state $\Gamma_{\mathrm{decay}}$ using the following
\begin{equation}
    \Gamma_{\mathrm{decay}} = 8 \alpha k |G_{1n}(k)|^2,
\label{eq:gamma_decay}
\end{equation}
where
\begin{equation}
    k = \sqrt{(BE_n(2) - BE_1(2))^2 - m_{\varphi}^2},
\end{equation}
and
\begin{equation}
    G_{1n}(k) = \int d^3r e^{-i \textbf{k} \cdot \textbf{r}/2} \psi_{n=1}^*(\textbf{r}) \psi_{n}(\textbf{r}).
\end{equation}
As long as $m_{\varphi} \ll m_{\chi} \alpha$, the potential describing the $\chi-\varphi$ interaction is approximately Coulombic, so the binding energies and wavefunctions $\psi$ are analogous to those of the Hydrogen atom for a Bohr radius of $a_0 = 2/ m_{\chi} \alpha$. Therefore, the binding energy is given by 
\begin{equation}
BE_n(2) = \frac{m_{\chi} \alpha^2}{4n^2}.
\label{eq:BE_twobody}
\end{equation}
Figure \ref{fig:BE_and_NumConst} shows the binding energy of the two-body bound state for various values of $m_\chi, \alpha_\chi$.

For realistic choices of $\alpha$ and $m_{\chi}$, we find that the rate for decay to the ground state exceeds the two-body formation rate $\Gamma_{\mathrm{decay}} \gg \Gamma_{\mathrm{form}}$. Here, $\Gamma_{\mathrm{form}}$ is calculated by assuming that the composites will form at a temperature of $T_{\mathrm{form}} = BE_1(2)/10$, then setting $\Gamma_{\mathrm{form}} = H(T_{\mathrm{form}})$, and $\Gamma_{\mathrm{decay}}$ is calculated using Equation \eqref{eq:gamma_decay}. Since the two-body composites decay quickly into their ground states after formation, Equations \eqref{eq:rho_chires_piecewise} and \eqref{eq:rho_phi_piecewise} can be simplified to
\begin{equation}
\rho_{\chi,\mathrm{res}}(a) =
\begin{cases}
    n_{\chi,\mathrm{res}}(a) m_{\chi} & a < a_{\mathrm{form}} \\ 
    n_{\chi,\mathrm{res}}(a) (m_{\chi}-BE_1(2)) & a \geq a_{\mathrm{form}},
\label{eq:rho_chires_piecewise2}
\end{cases}
\end{equation}
\begin{equation}
\rho_{\varphi}(a) =
\begin{cases}
    \frac{\pi^4 T}{30 \zeta(3)}n_{\varphi}(a) & a < a_{form} \\
    \frac{\pi^4 T}{30 \zeta(3)}n_{\varphi}(a) + \frac{a_{\mathrm{form}}^4 n_{\chi,\mathrm{res}}(a_{\mathrm{form}}) BE_1(2)}{a^4} & a \geq a_{\mathrm{form}}.
\label{eq:rho_phi_piecewise2}
\end{cases}
\end{equation}
Using these equations we can determine the relative energy density in $\chi$ versus $\varphi$ during two-body bound state formation. 

However, in this study we will find that the scalar field abundance $\rho_{\varphi}$ produced during N-body bound state formation greatly exceeds the abundance produced during formation of two-body bound states. 
This is because the binding energy of the large $N$-body composites greatly exceeds that of the two-body composites, meaning that the $\varphi$ energy density released from decay of $N$-body composites will greatly exceed that of two-body composites, cf. Equation \eqref{eq:BE_twobody} and Equation \eqref{eq:bindE}, which we will come to shortly.

\begin{figure}[H]
    \centering
    \includegraphics[width=0.6\columnwidth]{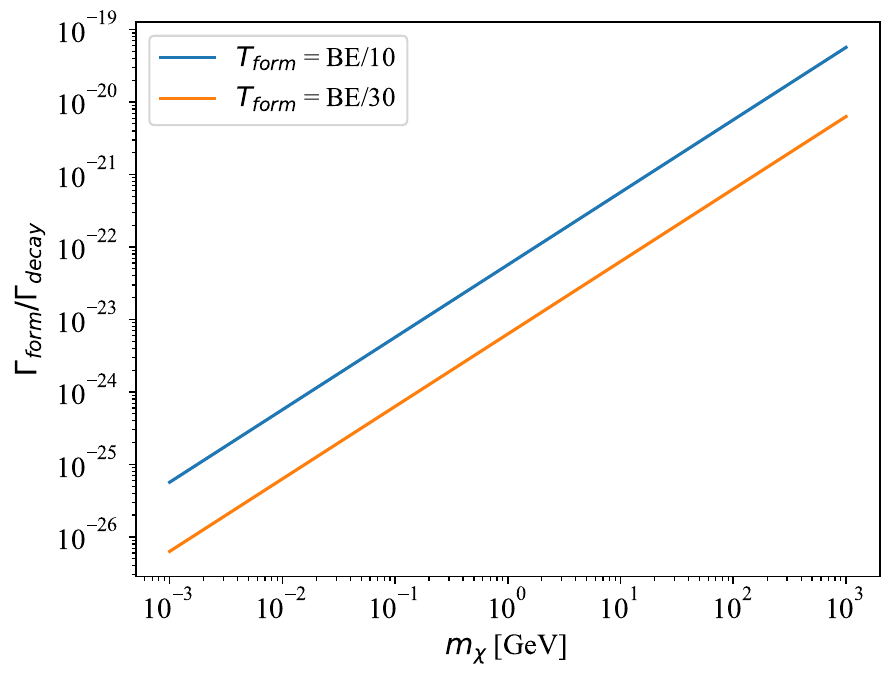}
    \caption{Ratio of formation rate to decay rate as a function of constituent mass $m_{\chi}$ for $\alpha$ = 0.1 and $N$ = 10, where $\Gamma_{\mathrm{form}} = H(T_{\mathrm{form}})$, and the decay rate is computed using Eq.~\eqref{eq:gamma_decay}. We see that $\Gamma_{form}/\Gamma_{decay} \ll 1$, justifying the assumption that the composite decays rapidly into its ground state for composites considered in this work.}
    \label{fig:gamma_ratio}
\end{figure}

Hence, we now turn directly to the formation and decay of large N-body bound states, and the $\varphi$ density that is emitted during this formation. After two-body states begin to form, larger composites form through aggregation and $\varphi$ emission ($X^N + X^N \rightarrow X^{2N} + \varphi$), with a rate $\Gamma \sim n_{X_N} \sigma_{X_N} v_{X_N}$. This $N$-body formation process eventually freezes out when $\Gamma/H \sim 1$, at the composite assembly freeze-out temperature defined as $T_{\mathrm{ca}}$.  Just as for the two-body composite, we approximate that the N-body composite forms at temperature $T_{\mathrm{form}} = BE_1(2)/10$, and we find that it decays quickly (relative to its formation time) into its ground state. Using the same procedure as for the two-body composites, we model the formation and decay of $N$-body bound states using Eqs.~\eqref{eq:rho_chires_piecewise2},\eqref{eq:rho_phi_piecewise2}, with the substitution $BE_1(2) \rightarrow BE_1(N)$, and we have checked that the instantaneous decay approximation holds by calculating the composite formation and decay rates for $N \gtrsim10$ for the case where the composite forms in its first excited state $n=2$. 
To do this, we use the same decay rate as for the two-body composite, but we define the Bohr radius in terms of the number of constituents in the composite as
\begin{equation}
    a_0 = \frac{2}{\alpha m_{\chi} N},
\end{equation}
and we define the aforementioned $N$-body binding energy as
\begin{equation}
    BE_n(N) = \frac{m_{\chi} \alpha^2 N^2}{4n^2}.
    \label{eq:bindE}
\end{equation}
In Figure \ref{fig:gamma_ratio}, we show the results of this calculation explicitly in the case where $\alpha$ = 0.1 and $N$ = 10.

We now turn to the structure and size of the composites. At the temperature of composite assembly, the expected number of constituents for a strongly-bound ($BE/N \simeq m_\chi$) composite is given by equating the N-composite formation rate with the Hubble rate, which yields the estimate \cite{Acevedo:2021kly,Hardy:2014mqa},
\begin{align}
    N &= \left(\frac{20 \sqrt{g_{\mathrm{ca}}^*} T_{\mathrm{m-r}} T_{\mathrm{ca}}^{3/2} M_{\mathrm{pl}}}{\bar{m}_{\chi}^{7/2} }\right)^{6/5}
    \simeq 10^{14} \left ( \frac{g_{\mathrm{ca}}^*}{10}\right )^{3/5} \left ( \frac{T_{\mathrm{ca}}}{1\, \mathrm{GeV}} \right )^{9/5} \left ( \frac{1 \, \mathrm{GeV}}{ \bar{m}_\chi} \right )^{21/5},
    \label{eq:num-const}
\end{align}
where $T_{\mathrm{m-r}}= 0.8$ eV is the matter-radiation equality temperature, and we reiterate that $\bar{m}_\chi$ is the effective mass of a constituent within the composite. Reference \cite{Gresham:2017cvl} has studied a number of composite formation prescriptions, and found that this is a reliable estimate for the number of constituents in an asymmetric fermionic composite.
The right panel of Figure \ref{fig:BE_and_NumConst} shows the number of constituents, in log scale, assuming that the binding energy per constituent is 0.99$m_\chi$ -- at the time that such a large composite forms, for a rapid composite decay rate relative to formation, most of the composite's unbound mass energy will have been liberated through $\varphi$ emission. In this figure, we also assume that composite assembly freezes out soon after formation begins, $T_{\mathrm{ca}} \approx T_{\mathrm{form}}$.

If the composites decay into their ground states quickly after forming, the energy transferred into $\varphi$ from formation and composite state decay corresponds to the binding energy of the ground state. For a given $\varphi$ mass, we calculate this binding energy using the relation $\Bar{m}_{\chi} = m_{\chi} - BE_1(N)$, where
\begin{equation}
    \Bar{m}_{\chi} = \left( \frac{3\pi m_{\chi}^2 m_{\varphi}^2}{2\alpha} \right)^{1/4}.
\label{eq:eff_mass}
\end{equation}
This last equation holds if the composite is saturated, which means that its number density and binding energy per constituent remain constant with increasing constituent number  \cite{Acevedo:2020avd}. The minimum number of constituents needed to reach saturation is $N_{\mathrm{sat}} \approx (r_0 m_\varphi)^{-3}$, where $r_0 = (\frac{9\pi}{4})^{1/3} (\bar{m}_\chi)^{-1}$, which is well-satisfied by all the parameter space we consider in this study.

In the top right and bottom plots of Figure \ref{fig:overall}, we show the cosmological energy density evolution of Standard Model radiation, dark matter fermions, and the scalar field for $m_{\varphi}$ = 1 meV, 1 eV, 100 eV, and for $m_{\chi} = 10$ GeV and $\alpha = 0.1$.

We see that for larger $m_{\varphi}$, the resulting energy density in $\varphi$ from composite formation and decay is smaller. The reason for this is somewhat subtle. First, we must note that the effective dark matter mass per constituent decreases from $m_\chi$ to $\bar m_\chi$ during composite assembly, and that $\bar m_\chi$ scales with $m_\varphi^{1/2}$ (Equation \eqref{eq:eff_mass}). This means that the initial abundance of $\chi_{\mathrm{res}}$ relative to the SM energy density, will need to be \emph{larger} for larger $m_\varphi$, for the resulting dark matter composite abundance to match the observed dark matter relic abundance. To put this another way, decreasing $m_{\varphi}$ results in a larger initial $\chi_{\mathrm{res}}$ abundance, which in turn yields a higher $\varphi$ abundance, after composite assembly.

\begin{figure}[H]
    \centering
    \includegraphics[width=\columnwidth]{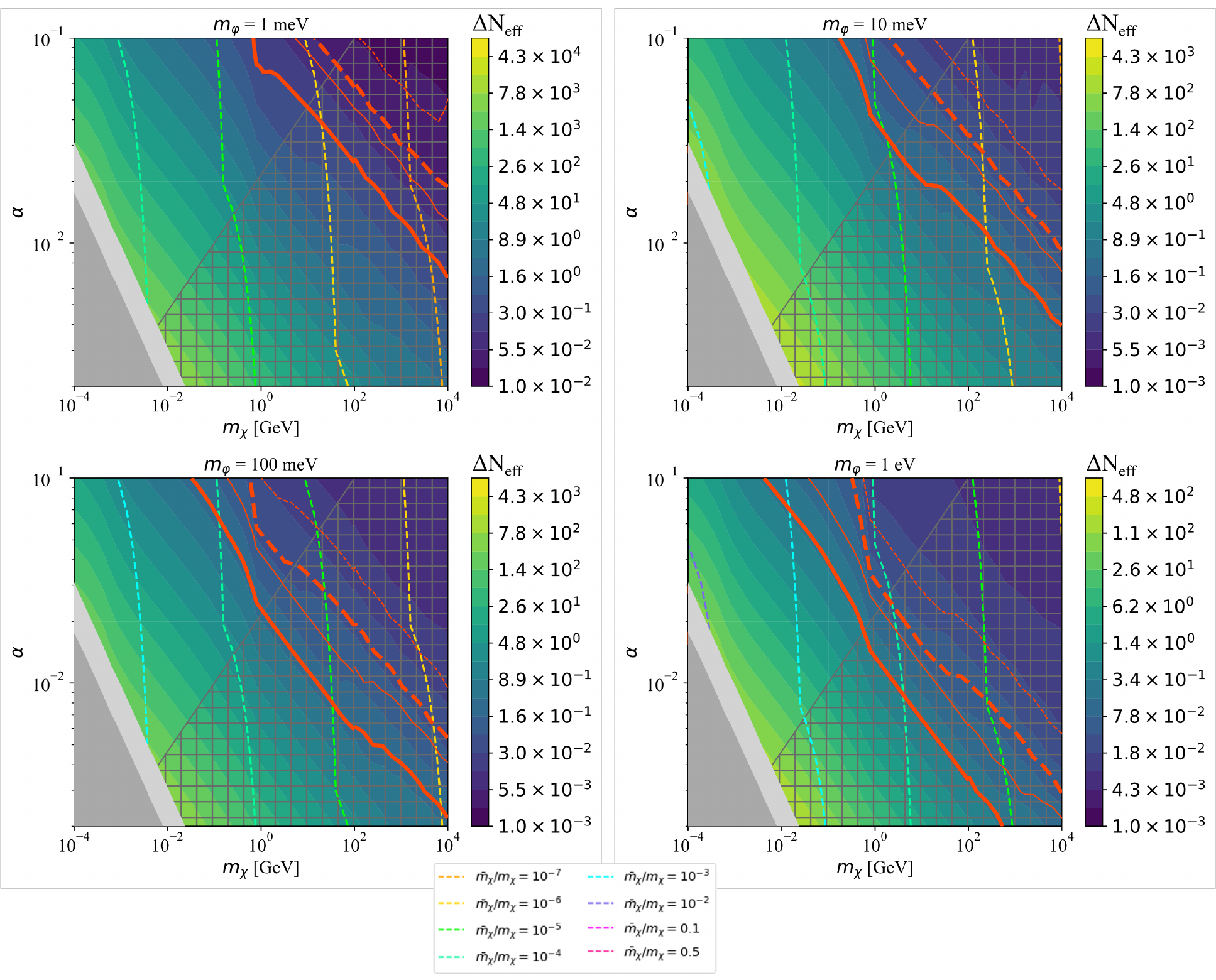}
    \caption{$\Delta N_{\mathrm{eff}}$ at $\chi$-SM equality in $m_{\chi}$-$\alpha$ parameter space for different indicated values of $m_{\varphi}$ between 1 meV and 1 eV, for N-body composites that decay into their ground state quickly after forming. The gridded dark grey region is where $\alpha$ is below the minimum value required to form composites  (absent some other attractive potential) and the solid grey region is where the composites form after matter-radiation equality. The coloured dashed lines show values for $\Bar{m}_{\chi}/m_{\chi}$, which increases as the ground state binding energy decreases. The region below the solid and dashed orange-red lines is excluded by 2$\sigma$ Planck bounds and by 2$\sigma$ projected CMB-S4 bounds on $N_{\mathrm{eff}}$ respectively. The thicker lines and darker grey region represent results for $T_{form}$ = BE/10, while the thinner lines and lighter grey region are for $T_{form}$ = BE/30; this range of formation temperatures corresponds to the uncertainty arising from Saha-type corrections to composite synthesis \cite{Gresham:2017cvl}.}
    \label{fig:Neff_vary_mphi}
\end{figure}

In Figure \ref{fig:Neff_vary_mphi}, we calculate the shift in relativistic degrees of freedom, $\Delta N_{\mathrm{eff}}$, at $\chi$-SM equality in $m_{\chi}$-$\alpha$ parameter space for fixed values of $m_{\varphi}$ ranging from 1 meV to 1 eV, where
\begin{equation}
    \Delta N_{\mathrm{eff}} = \frac{4g_{\varphi}}{7} \frac{\rho_{\varphi}}{\rho_{\mathrm{SM}}}.
\end{equation}
To compute $\Delta N_{\mathrm{eff}}$ we solve the Boltzmann Equations \eqref{eq:boltz_chi}-\eqref{eq:boltz_phi} to obtain the number density of each component, where the energy densities of $\chi_{\mathrm{res}}$ and $\varphi$ are given by the piecewise Equations \eqref{eq:rho_chires_piecewise2} and \eqref{eq:rho_phi_piecewise2} (with the aforementioned substitution $BE_1(2) \rightarrow BE_1(N)$). We compute the binding energy of the composites for a given $m_{\varphi}$ using Equation \eqref{eq:eff_mass}. Thus, we can place constraints on $m_{\chi}$ and $\alpha$ using the most recent 2$\sigma$ Planck bounds on $N_{\mathrm{eff}}$ \cite{Planck:2018vyg} as well as forecast the 2$\sigma$ projected bound on $N_{\mathrm{eff}}$ from CMB-S4, assuming 1' beams, 1 $\mu$K-arcmin temperature noise and a sky fraction of 0.5 \cite{CMB-S4:2016ple}. We also show how $\Bar{m}_{\chi}/m_{\chi}$, which increases as the ground state binding energy $BE_1(N)$ decreases, changes within this parameter space. 
We see that $\Delta N_{\mathrm{eff}}$ becomes smaller for larger $\alpha$ and larger $m_{\chi}$, because the composite formation time $T_{\text{form}} \simeq BE_1(2)/10$ is earlier, resulting in a smaller abundance of $\chi_{\mathrm{res}}$ relative to the Standard Model. There is some uncertainty in the estimation of $\Delta N_{\mathrm{eff}}$ from Saha-type corrections, associated with disassociation of the composites during multi-body assembly. It was found in reference \cite{Gresham:2017cvl} that this results in an effective composite synthesis temperature approximately between BE/10 and BE/30 for a range of masses and coupling strengths similar to those we consider here. Hence, we have also plotted the same $\Delta N_{\mathrm{eff}}$ contours for the case that $T_{\text{form}} \simeq BE_1(2)/30$, which provides some quantification of the uncertainty for the composite formation contribution to $\Delta N_{\mathrm{eff}}$.

For values of $m_{\varphi}$ larger than $\sim$1 eV, $N_{\mathrm{eff}}$ bounds are no longer applicable since $\varphi$ becomes matter-like. However, the parameter space is still constrained by requiring that matter-radiation equality occur around $T = 0.8$ eV, $i.e.$ that the $\varphi$ and composite abundance together do not exceed the observed abundance of dark matter. The energy density evolution of the Standard Model for a case like this is also shown in the bottom right plot of Figure \ref{fig:overall}.

\begin{figure}[H]
    \centering
    \includegraphics[width=\columnwidth]{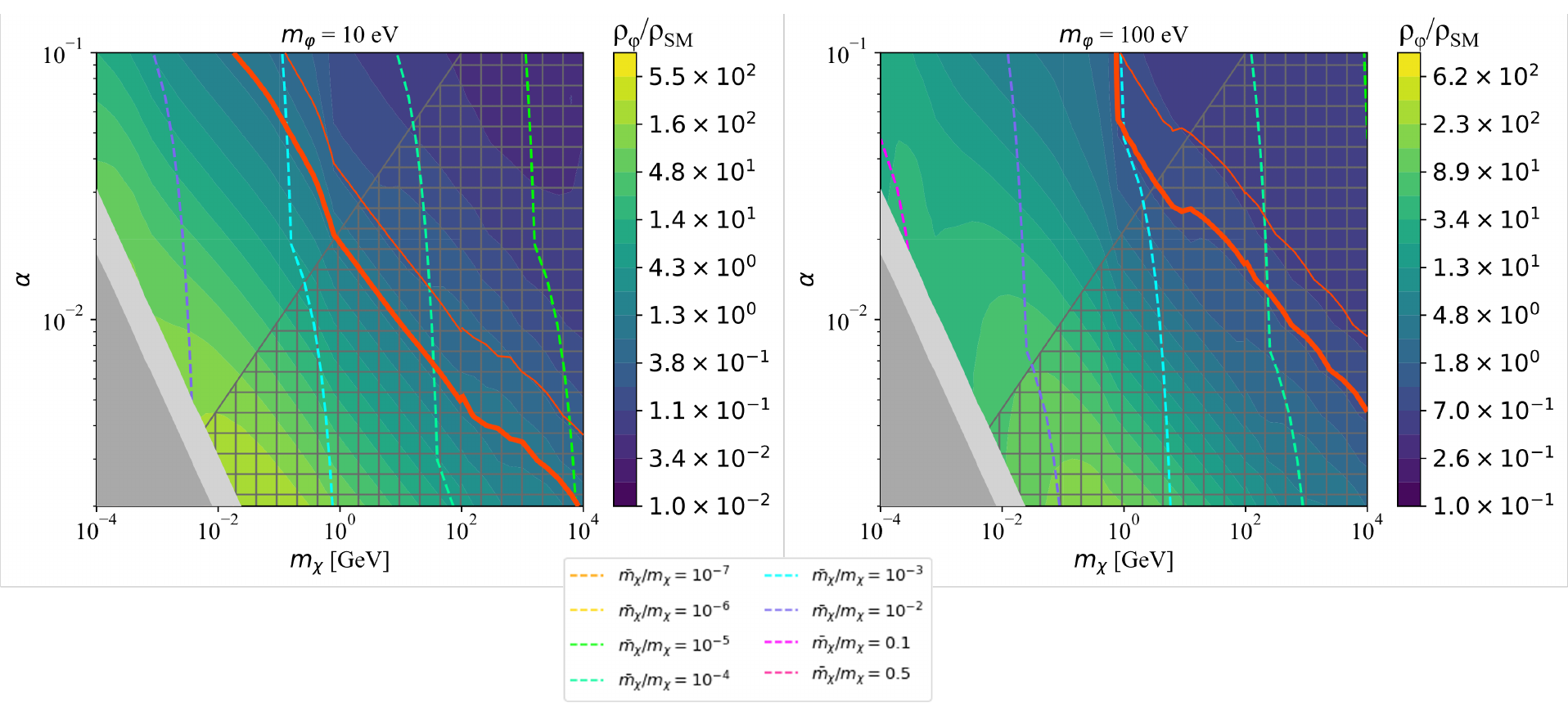}
    \caption{The same as Figure \ref{fig:Neff_vary_mphi}, but for heavier $m_\varphi =10,\,100$ eV. Here we show the relative $\varphi-SM$ density, $\rho_{\varphi}/\rho_{SM}$ at matter-radiation ($i.e.$ $\chi$-SM) equality. Here again it is assumed that the N-body composite decays into its ground state immediately after forming. The gridded dark grey region is where $\alpha$ is below the minimum value required to form composites, and the solid grey region is where the composite would form and decay after matter-radiation equality. The region below the orange-red line is where $\rho_{\varphi}/\rho_{SM}$ $\geq$ 1, which is excluded by requiring matter-radiation equality occur at $T\sim$ eV. The thicker lines and darker grey region represent results for $T_{form}$ = BE/10, while the thinner lines and lighter grey region are for $T_{form}$ = BE/30; this range of formation temperatures corresponds to the uncertainty arising from Saha-type corrections to composite synthesis.}
    \label{fig:rhophiSM_vary_mphi}
\end{figure}

In Figure \ref{fig:rhophiSM_vary_mphi}, we show the ratio $\rho_{\varphi}/\rho_{\mathrm{SM}}$ at $\chi$-SM equality in $m_{\chi}$-$\alpha$ parameter space for $m_{\varphi}$ = 10 eV and for $m_{\varphi}$ = 100 eV. We can calculate the energy density fraction using the same equations as for $N_{\mathrm{eff}}$. However, to take into account $\varphi$ becoming matter-like as the temperature of the Universe decreases, we generalize Equation \eqref{eq:rho_phi_piecewise2} to
\begin{equation}
\rho_{\varphi}(a) =
\begin{cases}
    E(a)n_{\varphi}(a) & a < a_{\mathrm{form}} \\
    E(a)n_{\varphi}(a) + \frac{E(a)n_{\varphi}(a)}{E(a_{\mathrm{form}})n_{\varphi}(a_{\mathrm{form}})} n_{\chi,\mathrm{res}}(a_{\mathrm{form}}) BE_1(2)  & a \geq a_{\mathrm{form}},
\end{cases}
\end{equation}
where $E = \gamma m_{\varphi}$, $\gamma = \sqrt{\frac{1}{1-v^2}}$, and $v = \frac{2(T/m_{\varphi})(T/m_{\varphi}+1)e^{-m_{\varphi}/T}}{K_2 (m_{\varphi}/T)}$ is the average speed defined by the Maxwell-J\"uttner distribution.

Altogether, we see that secluded composites of the kind we have studied predict a nontrivial relic abundance of light scalar particles. These could be discovered independently of the dark composites, either through cosmological observations like $\Delta N_{\mathrm{eff}}$, or perhaps through direct observation of the $\varphi$ state. Some details will depend on how $\varphi$ and $\chi$ couple to the Standard Model, as we investigate in the next Section.

\section{High temperature and Higgs portal couplings}\label{sec:3}

\begin{figure}
    \centering
    \includegraphics[width=0.85\linewidth]{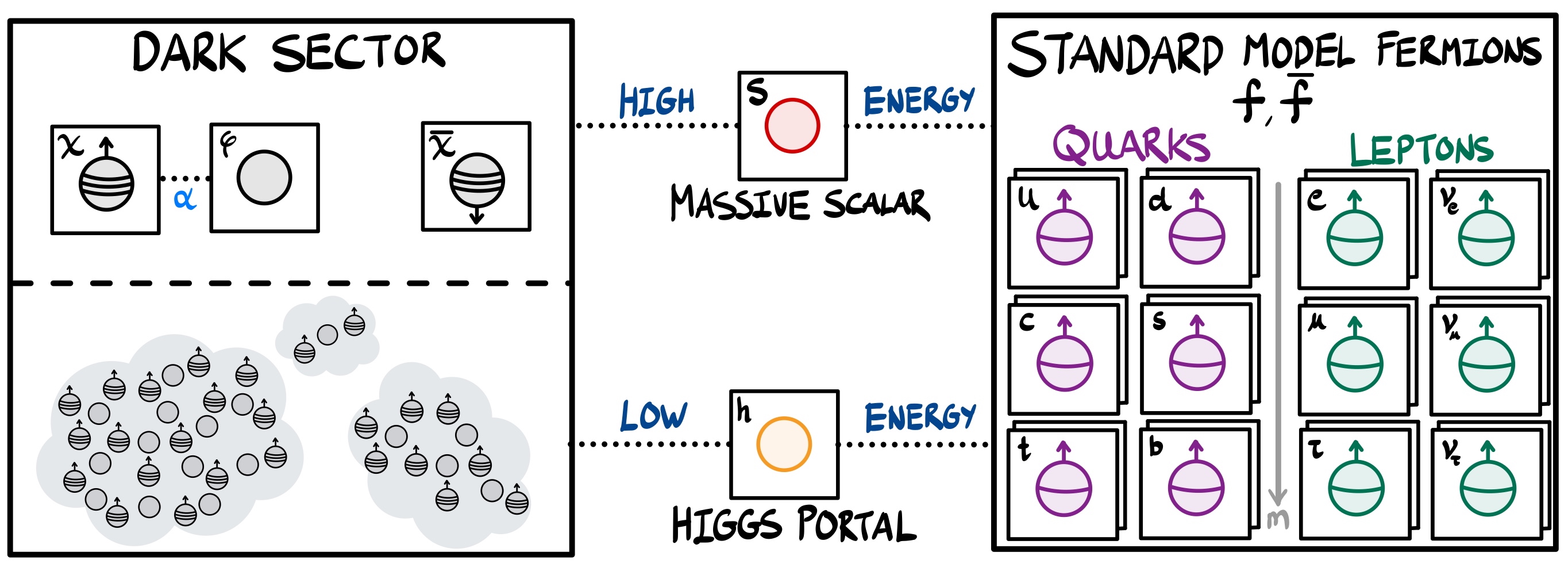}
    \caption{Schematic illustrating the interactions between DM and SM in our model. In the minimal model explored in this paper, the composite dark matter sector ($\chi, \varphi$) equilibrates at high temperature through a massive $m_{S} \gg 10^7 $ GeV scalar $S$ that couples also to the Standard Model. A small Higgs portal coupling between $\varphi$ and $H$ (possibly generated by loops) leads to detection prospects for the composites via an effective Yukawa coupling to nucleons \cite{Acevedo:2020avd,Acevedo:2021kly}.}
    \label{fig:enter-label}
\end{figure}

So far we have explored the formation of composite dark matter states in a dark matter model comprised of asymmetric dark matter (fermions $\chi$) and a light scalar ($\varphi$), which results in dark matter composites. This kind of dark matter can lead to many phenomenological possibilities that depend on the composite's couplings to the Standard Model \cite{Gresham:2018anj,Acevedo:2020avd,Acevedo:2021kly}. This is the focus of this section, where we study a UV completion of the model and a Higgs portal coupling to $\varphi$ at low temperatures. We will also consider couplings for the dark matter to equilibrate at high temperatures before becoming secluded, as was assumed in Section \ref{sec:2}. These interactions between dark matter and the Standard Model are summarized schematically in Figure \ref{fig:enter-label}.

In Subsection \ref{sec:UV} we detail the UV completion of our DM model vis-a-vis its cosmological abundance, and particularly the mechanism for the secluded dark sector to equilibrate with the SM at high temperatures. This involves a heavy scalar mediator that couples to both $\chi$ and the SM, and thus facilitates interactions between the dark fermions and SM fermions at high temperatures, while falling out of equilibrium at lower temperatures. 

Following this, in Subsection \ref{sec:higgs} we introduce a Higgs portal interaction, mixing $\varphi$ with the SM Higgs, and detail its resultant interactions after the electroweak phase transition (EWPT). This Higgs portal implies potential detection signatures. In particular, we explore the implications for direct detection, stellar cooling, and fifth force tests, while also determining how cosmological considerations restrict the Higgs portal couplings.

\subsection{High Energy Thermal Coupling}
\label{sec:UV}

During the freeze-out and composite assembly phases, we assumed the dynamics of our composite dark matter model were governed by the interactions present in Equation \eqref{eq:lagrangian_ds} (excluding the final term), and that the dark sector did not have appreciable SM interactions. In addition, we set the initial density of the dark sector by assuming that at high temperatures it was in equilibrium with the SM. We now lay out a simple high-temperature model that justifies this assumption. An early universe model ($i.e.$ $T > T_{\text{EWPT}}$) along these lines consists of a heavy scalar field $S$ that couples to both the dark fermions $\chi$ and SM fermions $f$. The Lagrangian is then
\begin{align}
    \mathcal{L}_S = \frac{1}{2}\left(\partial_{\mu} S \right)^2+ \frac{1}{2} m_S^2 S^2 + g_{S\chi} S \bar{\chi}\chi + g_{S f} S \bar{f}f 
\end{align}
where both factors of $g_S$ and $m_S$ are free parameters. At temperatures $T$, if we have $m_S\gg T$, then we can integrate out the heavy scalar $S$ to get,
\begin{align}
    \mathcal{L}_S^{\mathrm{int}} \supset \frac{2 g_{S\chi} g_{Sf}}{m_S^2} \bar{\chi}\bar{f}f \chi
\end{align}

Such a high energy thermal coupling equilibrates the dark sector abundance at some high temperature $T_{RH} \gg T_{\mathrm{EWPT}}$, and then later decouples at some temperature $T_\mathrm{dec}$, where $T_{\mathrm{RH}}$ corresponds to some early time reheating temperature. Assuming $g_{S\chi}g_{Sf} \sim \mathcal{O}(1)$, the cross section for the equilibration interaction is
\begin{align}
   \langle \sigma v \rangle \sim \left( \frac{1}{4\pi}  \right)^3 \left(\frac{2}{m_S} \right)^2 = \frac{1}{32\pi^3 m_S^2}. \label{eq:uv_cross_section}
\end{align}
For the dark sector to be secluded from the SM during composite assembly, we require the UV decoupling $T_\mathrm{dec}$ temperature to be higher than the freeze-out temperature for the dark matter particle $m_\chi$. This implies that $T_{\mathrm{dec}}\gtrsim m_\chi/10$. Combining this with the observed value of the dark matter density and Equation \eqref{eq:uv_cross_section}, we can derive a bound on $m_S$
\begin{align}
    m_S > 10^7 \, \mathrm{GeV} \left(\frac{m_\chi}{1\, \mathrm{GeV}}\right)^{\frac{1}{2}}.
\end{align}
For a similar discussion of decoupled dark sectors see $e.g.$ \cite{Elahi:2014fsa}. Having detailed a model for early universe equilibration between the SM and the dark sector, we now turn to the low energy coupling that may arise between these sectors.

\subsection{Higgs Portal} \label{sec:higgs}

To investigate the low energy coupling between the SM and a secluded dark sector, we will specifically consider a Higgs portal model where the effective couplings are defined by the following Lagrangian,
\begin{align}
	\mathcal{L}_{\text{Higgs-portal}} = \begin{cases}\lambda_{h\varphi}h^2(\varphi + v_{\varphi})^2& T>T_{\mathrm{EWPT}}\\
	\lambda_{h \varphi}\left( h + v_{h} \right) ^2(\varphi + v_{\varphi})^2&T_{\mathrm{EWPT}}>T \label{eq:higgs_portal}
    \end{cases} 
\end{align}
As is often the case for Higgs portal models \cite{Silveira:1985rk,Heeba:2018wtf}, we assume the dark scalar field obtains a vacuum expectation value (vev), $ \varphi \to \varphi + v_{\varphi}$, before EWPT. After EWPT, we have the Higgs field vev $ h \to h + v_{h} $, where $h$ is the low energy SM Higgs field. The mixing between the Higgs field and the scalar is given by the mixing angle $\sin \theta=(\lambda_{h\varphi}v_h v_\varphi)/(m_h^2-m_\varphi^2)$. 

This Higgs portal interaction potentially leads to a freeze-in contribution of $\varphi$ particles from interactions with the SM bath. 
The relevant Lagrangian term that contributes to processes whereby SM fields produce $\varphi$ in the thermal bath, is given by $2 v_h \lambda_{h\varphi} h \varphi^2$ in Equation \eqref{eq:higgs_portal}. This term results in $h \to \varphi \varphi$ for light $\varphi$. It has been shown that so long as $\lambda_{h\varphi} < 4 \times 10^{-8}$, $\varphi$, assuming a standard radiation-dominated cosmology and $m_\varphi \ll m_h$, this channel produces a $\varphi$ abundance smaller than the dark matter density \cite{Lebedev:2021xey}.

\begin{figure}[H]
    \centering
    \includegraphics[width=\linewidth]{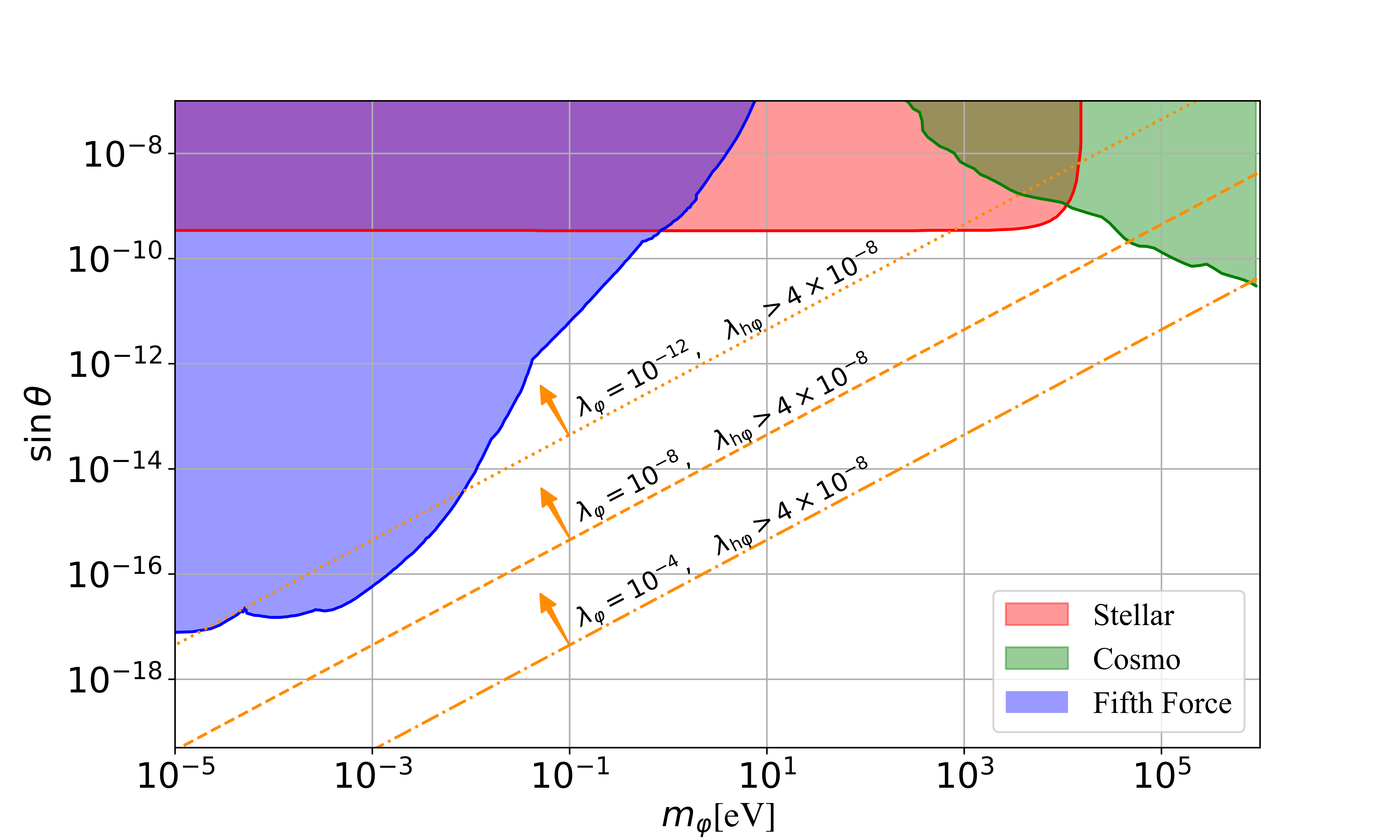}
    \caption{ Exclusion regions for the mixing angle $\theta$ as a function of the scalar mass $m_\varphi$ in a Higgs portal scalar model. Shaded regions represent constraints from stellar observations (red) and cosmology (green) (both from \cite{Lasenby}), and fifth-force experiments (blue). The stellar cooling bounds are comparable to the white dwarf bounds obtained recently in \cite{Bottaro:2023gep}. The fifth-force bounds were derived by recasting limits on ${m_\varphi, \alpha_{\mathrm{grav}}}$ in \cite{ff_constraint_1,ff_constraint_2,ff_constraint_3,ff_constraint_4} in terms of $\sin\theta$ via Equation \eqref{eq:theta_to_alpha}. The orange dotted lines correspond to the saturation of the condition $\lambda_{h\varphi} < 4 \times 10^{-8}$, above these lines (for indicated values of $\lambda_\phi$) the $\varphi$ abundance from $h \to \varphi \varphi$ exceeds the dark matter abundance. The multiplicity of these lines arises because $\theta$ and $\lambda_{h\varphi}$ are degenerate without a specified scalar self-coupling $\lambda_\varphi$: using $v_\varphi = m_\varphi / \sqrt{2\lambda_\varphi}$ and assuming $m_\varphi \ll m_h$, the mixing angle scales as $\sin \theta \approx \lambda_{h\varphi} v_h m_\varphi / (m_h^2 \sqrt{2\lambda_\varphi})$. We note that the cosmological constraints shown in green originate from electromagnetic energy injection in the post-recombination Universe: decays occurring at timescales between approximately $10^6$ s and $10^9$ s lead to $\mu$ and $y$-type spectral distortions in the CMB, while later decays produce photons that survive to presence day and either contribute to or distort the extragalactic background light (EBL) spectrum \cite{Flacke:2016szy,Hardy:2016kme}. These bounds are effective up to scalar masses of order $m_\varphi\sim10^6 \rm eV$. Further breakdown of cosmological constraints on Higgs Portal scalars is found in \cite{Fradette:2018hhl}. For heavier scalars and larger mixing parameter (not shown in figure), additional constraints arise from Big Bang Nucleosynthesis, where late-time decays can disrupt nuclear process or inject entropy, thereby affecting the baryon-photon ratio. These limits  exclude mixing angles up to $\sin \theta \sim 10^{-5}$ and extend to masses as large as $m_\varphi \sim \rm 200$ MeV. For even larger mixing angles, $\sin\theta \gtrsim 10^{-4} $, collider searches - in particular Kaon decays - impose additional exclusions \cite{Goudzovski:2022vbt}.   As a result, there remains another narrow window—less than one order of magnitude in $\sin\theta$—between the upper edge of the BBN bounds and the lower edge of the collider bounds where the model is still viable.}
    \label{fig:constraints}
\end{figure}

\paragraph{Direct Detection of Secluded Composite Dark Matter}

The Higgs portal coupling implies a spin-independent elastic scattering process, from the dark composites coupling to nucleons via a dark scalar \cite{Acevedo:2020avd,Acevedo:2021kly}. We can define this effective $\varphi$-Nucleon coupling as follows, 
\begin{align}
    \mathcal{L}_{\varphi N}^{\mathrm{(eff)}} = \frac{\lambda_{h\varphi} m_N f_N}{2 m_{h}^2} \varphi^2 \bar{N}N \label{eq:higgs_eff} , \qquad \mathrm{where} \quad f_N = \sum_{q \in \{u,d,s\}} f_q + \frac{2}{9} f_{TG} \simeq 0.3,
\end{align} 
where the parameters $f_q$ and $f_{TG}$ correspond to the contribution from nucleon interaction with quarks and gluons respectively \cite{Ellis:2008hf}. Using this interaction, we have a spin-independent $\varphi$-DM scattering cross section given by
\begin{align}
    \sigma_{\varphi N}^{\text{SI}} = \frac{\lambda_{h \varphi}^2 f_N^2 }{ 4 \pi m_h^4} \frac{m_N^4}{(m_\varphi+m_N)^2} \approx 10^{-53} \left( \frac{\lambda_{h\varphi}}{4 \times 10^{-8}} \right)\, \text{cm}^2, \quad \text{for } m_\varphi \ll m_N.
\end{align}
At present, ferreting out such low cross sections for scalar dark matter ($\varphi$) at a direct detection experiment is impractical. 

On the other hand, Refs.~\cite{Acevedo:2020avd,Acevedo:2021kly} showed that the scalar field binding a fermionic composite together can generate substantial interactions between the composite state and nuclei through a nucleonic Yukawa potential term of the form $g_N \bar N \varphi N$. This is because Standard Model nuclei can recoil off (or be accelerated by) the internal potential of the dark matter composite. In terms of the Higgs portal scalar interaction this Yukawa term is defined as
\begin{align}
    \mathcal{L}_N^{\text{(Yukawa)}} = \frac{\lambda_{h\varphi} m_N f_N \langle \varphi \rangle}{2 m_h^2}  \bar{N} \varphi N = g_N \bar{N} \varphi N, \label{eq:nucleon_yukawa}
\end{align}
where $\langle \varphi \rangle = m_\varphi/\sqrt{2 \lambda_\varphi}$ is the scalar's vev in vacuum, but this vev can reach much larger values inside a fermionic dark matter composite.
We now briefly consider elastic scattering of nuclei by fermionic composite DM at direct detection experiments such as Xenon1T, PandaX, and LZ, where detection prospects have been demonstrated for values of $g_N$ as low as $10^{-17}$ \cite{Acevedo:2021kly}. In terms of the Higgs portal coupling parameters we find
\begin{align}
    g_N = 10^{-16} \left(\frac{\lambda_{h\varphi}}{10^{-9}}\right)  \left(\frac{\langle \varphi \rangle}{\text{GeV}} \right),
\end{align}
where parametrically $\langle \varphi \rangle \sim m_\chi / g_{\chi \varphi}$ inside a composite state with a large number of fermions $\chi$ \cite{Acevedo:2021kly}.
From this we see that the secluded composite scenario is detectable through low energy nuclear recoils at sufficiently advanced direct detection experiments and also potentially through other high energy processes at neutrino experiments \cite{Acevedo:2020avd}. 

\paragraph{Fifth Force Constraints on $\varphi$ Coupling to Nucleons}

Next we examine the limits on a Higgs-portal dark scalar from equivalence principle ($aka$ fifth-force) tests for the Yukawa potential 
\begin{align}
V(r) =  \alpha_N G m_N^2\frac{ e^{-m_\varphi r}}{r},
\end{align}
where this potential is normalized according to conventions used in torsion pendulum fifth-force experiments; $G$ is Newton's constant and $m_N$ is the nucleon mass. Matching this to the Higgs portal Yukawa term in Equation \eqref{eq:nucleon_yukawa} we find
\begin{align}
\alpha_N = \frac{1}{32 \pi G} \frac{\lambda_{h\varphi}^2f_N^2 m_{\varphi}^2}{m_h^4 \lambda_\varphi} ,
\end{align}
in terms of Higgs portal coupling parameters. Recasting this expression in terms of the Higgs portal mixing angle $\theta$ yields
\begin{align}
\sin \theta = \frac{\sqrt{16 \pi G \alpha_N} v_h}{f_N} \label{eq:theta_to_alpha}.
\end{align}
We use this expression to recast \cite{ff_constraint_1,ff_constraint_2,ff_constraint_3,ff_constraint_4} and obtain fifth-force bounds shown in Figure \ref{fig:constraints}.


\paragraph{Scalar Decay}
We have explored dark scalars with masses in the range $m_\varphi \in [\rm {\mu}\text{eV}, \text{eV}]$. Within this range, the only kinematically allowed decay channel is to photons, resulting in rather long scalar lifetimes, see $e.g.$ \cite{Bramante:2016yju}. Such decays are absent at the tree-level but can be induced by loops of SM fermions and bosons. The partial width is given by \cite{Heeba:2018wtf},
\begin{align}
    \Gamma(\varphi \to \gamma \gamma) = \sin^2 \theta \frac{\alpha_e^2 m_\varphi^3}{256 \pi^3 v_h^2} \left| f\left(z_f\right) + \frac{7}{3} \right|^2
    \label{eq:decay_to_photon}
\end{align}
where $z_f = 4 m_f^2/m_\varphi^2$ for fermion mass $m_f$, $\alpha_e$ is the fine structure constant, and $f$ is a form factor for fermionic loops given by,
\begin{align}
    f(\tau) = 2 \tau \left[ 1 + (1- \tau) \arctan^2\left(\frac{1}{\sqrt{\tau - 1}}\right) \right]
\end{align}
The second term in this factor only contributes for leptonic loops. For all the other quarks, we can take $z_f \to \infty$, and then we would be left with just the first term. We can see from this that in the parameter space of interest, we should expect the Higgs portal scalars we have studied to be cosmologically long-lived. Inverting Equation \eqref{eq:decay_to_photon} yields
\begin{align}
    \tau_{\varphi \rightarrow \gamma \gamma} \approx 7.58 \times 10^{43}  \, \text{s} \left( \frac{4 \times 10^{-8}}{\lambda_{h\varphi}} \right)^2 \left( \frac{1 \, \text{eV}}{m_\varphi} \right)^5 \left(\frac{\lambda_\varphi}{10^{-8}}\right),
\end{align}
so for parameters of interest in this work, the scalar field $\varphi$ is cosmologically long-lived.

Figure \ref{fig:constraints} shows parameter space for a Higgs-portal coupling to secluded asymmetric dark matter composites, along with relevant constraints from fifth-force experiments, stellar cooling, and cosmological observations.

\section{Conclusion}\label{sec:4}
In this article, we explored a secluded composite dark matter model where dark matter particles are made up of fermions $\chi$ bound together by a light scalar field $\varphi$, and the dark sector forms composite states out of equilibrium with the Standard Model. 

One goal of this work was to determine how different processes, like $\chi$ annihilation and composite formation, contribute to the energy density of $\varphi$ in the early Universe. We found that while the formation of two-body states does not significantly change the energy density of $\varphi$, for larger N-body bound states, the energy density of $\varphi$ can be augmented appreciably, leading to observable cosmological effects. Our results indicate that for scalar masses ranging from $100\,\mu\mathrm{eV}$ to $1 \, \mathrm{eV} $, there is parameter space that is ruled out by Planck observations, and also discoverable using future cosmological observations like CMB-S4. For scalar masses above an eV, $\varphi$ would serve as a subcomponent of dark matter, leading to a separate class of observable signatures, depending on $\varphi$'s coupling to the Standard Model. We leave the research into detecting a $\varphi$ subcomponent of dark matter to future work.

To further determine the dynamics of secluded composite dark matter, we introduced a Higgs portal between $\varphi$ and Standard Model particles. 
Such a portal coupling implies scattering and other interactions between composites and SM nuclei, investigated previously in \cite{Acevedo:2020avd,Acevedo:2021kly}. Here we have explored how an explicit Higgs portal UV completion is limited by  astrophysical constraints, fifth-force tests, model consistency, and cosmological limits on the freeze-in production of $\varphi$.

In future work, the secluded composite assembly scenario could be extended to explore different cosmological interactions between the dark and visible sectors, that would affect composite formation and abundance. In particular, this work assumed equal temperatures for the dark and visible sectors, stemming from an early period of equilibration. More generally, these two sectors might evolve at different temperatures. It would also be useful to investigate cases where freeze-out and composite assembly happen at drastically different temperatures, and perhaps where composite assembly occurs even after matter-radiation equality, which would possibly lead to distinct observational signatures in cosmological parameters like  $\Delta N_{\mathrm{eff}} $.

In summary, we have examined the impact of light scalar fields in composite dark matter dynamics and shown how a secluded dark sector could leave detectable imprints on cosmological observables, and how a small Higgs portal coupling between composite dark matter and SM fields can lead to detectable interactions.

\bibliographystyle{JHEP.bst}

\bibliography{ccosmo.bib}

@article{Elahi:2014fsa,
    author = "Elahi, Fatemeh and Kolda, Christopher and Unwin, James",
    title = "{UltraViolet Freeze-in}",
    eprint = "1410.6157",
    archivePrefix = "arXiv",
    primaryClass = "hep-ph",
    doi = "10.1007/JHEP03(2015)048",
    journal = "JHEP",
    volume = "03",
    pages = "048",
    year = "2015"
}

@PREAMBLE{
 "\providecommand{\noopsort}[1]{}" 
 # "\providecommand{\singleletter}[1]{#1}%" 
}

@article{Acevedo:2021kly,
    author = "Acevedo, Javier F. and Bramante, Joseph and Goodman, Alan",
    title = "{Accelerating composite dark matter discovery with nuclear recoils and the Migdal effect}",
    eprint = "2108.10889",
    archivePrefix = "arXiv",
    primaryClass = "hep-ph",
    doi = "10.1103/PhysRevD.105.023012",
    journal = "Phys. Rev. D",
    volume = "105",
    number = "2",
    pages = "023012",
    year = "2022"
}

@article{Acevedo:2020avd,
    author = "Acevedo, Javier F. and Bramante, Joseph and Goodman, Alan",
    title = "{Nuclear fusion inside dark matter}",
    eprint = "2012.10998",
    archivePrefix = "arXiv",
    primaryClass = "hep-ph",
    doi = "10.1103/PhysRevD.103.123022",
    journal = "Phys. Rev. D",
    volume = "103",
    number = "12",
    pages = "123022",
    year = "2021"
}

@article{Coskuner:2018are,
    author = "Coskuner, Ahmet and Grabowska, Dorota M. and Knapen, Simon and Zurek, Kathryn M.",
    title = "{Direct Detection of Bound States of Asymmetric Dark Matter}",
    eprint = "1812.07573",
    archivePrefix = "arXiv",
    primaryClass = "hep-ph",
    doi = "10.1103/PhysRevD.100.035025",
    journal = "Phys. Rev. D",
    volume = "100",
    number = "3",
    pages = "035025",
    year = "2019"
}

@article{Bramante:2018tos,
    author = "Bramante, Joseph and Broerman, Benjamin and Kumar, Jason and Lang, Rafael F. and Pospelov, Maxim and Raj, Nirmal",
    title = "{Foraging for dark matter in large volume liquid scintillator neutrino detectors with multiscatter events}",
    eprint = "1812.09325",
    archivePrefix = "arXiv",
    primaryClass = "hep-ph",
    doi = "10.1103/PhysRevD.99.083010",
    journal = "Phys. Rev. D",
    volume = "99",
    number = "8",
    pages = "083010",
    year = "2019"
}

@article{Digman:2019wdm,
    author = "Digman, Matthew C. and Cappiello, Christopher V. and Beacom, John F. and Hirata, Christopher M. and Peter, Annika H.G.",
    title = "{Not as big as a barn: Upper bounds on dark matter-nucleus cross sections}",
    eprint = "1907.10618",
    archivePrefix = "arXiv",
    primaryClass = "hep-ph",
    doi = "10.1103/PhysRevD.100.063013",
    journal = "Phys. Rev. D",
    volume = "100",
    number = "6",
    pages = "063013",
    year = "2019"
}

@article{Jacobs:2014yca,
    author = "Jacobs, David M. and Starkman, Glenn D. and Lynn, Bryan W.",
    title = "{Macro Dark Matter}",
    eprint = "1410.2236",
    archivePrefix = "arXiv",
    primaryClass = "astro-ph.CO",
    doi = "10.1093/mnras/stv774",
    journal = "Mon. Not. Roy. Astron. Soc.",
    volume = "450",
    number = "4",
    pages = "3418--3430",
    year = "2015"
}

@article{Fedderke:2024hfy,
    author = "Fedderke, Michael A. and Kaplan, David E. and Mathur, Anubhav and Rajendran, Surjeet and Tanin, Erwin H.",
    title = "{Fireball antinucleosynthesis}",
    eprint = "2402.15581",
    archivePrefix = "arXiv",
    primaryClass = "hep-ph",
    doi = "10.1103/PhysRevD.109.123028",
    journal = "Phys. Rev. D",
    volume = "109",
    number = "12",
    pages = "12",
    year = "2024"
}

@article{Bramante:2018qbc,
      author         = "Bramante, Joseph and Broerman, Benjamin and Lang, Rafael
                        F. and Raj, Nirmal",
      title          = "{Saturated Overburden Scattering and the Multiscatter
                        Frontier: Discovering Dark Matter at the Planck Mass and
                        Beyond}",
      journal        = "Phys. Rev.",
      volume         = "D98",
      year           = "2018",
      number         = "8",
      pages          = "083516",
      doi            = "10.1103/PhysRevD.98.083516",
      eprint         = "1803.08044",
      archivePrefix  = "arXiv",
      primaryClass   = "hep-ph",
      SLACcitation   = "%%CITATION = ARXIV:1803.08044;%%"
}

@article{Bhoonah:2020dzs,
    author = "Bhoonah, Amit and Bramante, Joseph and Schon, Sarah and Song, Ningqiang",
    title = "{Detecting Composite Dark Matter with Long Range and Contact Interactions in Gas Clouds}",
    eprint = "2010.07240",
    archivePrefix = "arXiv",
    primaryClass = "hep-ph",
    month = "10",
    year = "2020"
}

@article{Dhakal:2022rwn,
    author = "Dhakal, Pawan and Prohira, Steven and Cappiello, Christopher V. and Beacom, John F. and Palo, Scott and Marino, John",
    title = "{New constraints on macroscopic dark matter using radar meteor detectors}",
    eprint = "2209.07690",
    archivePrefix = "arXiv",
    primaryClass = "hep-ph",
    doi = "10.1103/PhysRevD.107.043026",
    journal = "Phys. Rev. D",
    volume = "107",
    number = "4",
    pages = "043026",
    year = "2023"
}

@article{Clark:2020mna,
    author = "Clark, Michael and Depoian, Amanda and Elshimy, Bahaa and Kopec, Abigail and Lang, Rafael F. and Qin, Juehang",
    title = "{Direct Detection Limits on Heavy Dark Matter}",
    eprint = "2009.07909",
    archivePrefix = "arXiv",
    primaryClass = "hep-ph",
    month = "9",
    year = "2020"
}

@article{Grabowska:2018lnd,
    author = "Grabowska, Dorota M. and Melia, Tom and Rajendran, Surjeet",
    title = "{Detecting Dark Blobs}",
    eprint = "1807.03788",
    archivePrefix = "arXiv",
    primaryClass = "hep-ph",
    doi = "10.1103/PhysRevD.98.115020",
    journal = "Phys. Rev. D",
    volume = "98",
    number = "11",
    pages = "115020",
    year = "2018"
}

@article{Laha:2013gva,
    author = "Laha, Ranjan and Braaten, Eric",
    title = "{Direct detection of dark matter in universal bound states}",
    eprint = "1311.6386",
    archivePrefix = "arXiv",
    primaryClass = "hep-ph",
    doi = "10.1103/PhysRevD.89.103510",
    journal = "Phys. Rev. D",
    volume = "89",
    number = "10",
    pages = "103510",
    year = "2014"
}

@article{Laha:2015yoa,
    author = "Laha, Ranjan",
    title = "{Directional detection of dark matter in universal bound states}",
    eprint = "1505.02772",
    archivePrefix = "arXiv",
    primaryClass = "hep-ph",
    doi = "10.1103/PhysRevD.92.083509",
    journal = "Phys. Rev. D",
    volume = "92",
    pages = "083509",
    year = "2015"
}

@article{Hardy:2014mqa,
      author         = "Hardy, Edward and Lasenby, Robert and March-Russell, John
                        and West, Stephen M.",
      title          = "{Big Bang Synthesis of Nuclear Dark Matter}",
      journal        = "JHEP",
      volume         = "06",
      year           = "2015",
      pages          = "011",
      doi            = "10.1007/JHEP06(2015)011",
      eprint         = "1411.3739",
      archivePrefix  = "arXiv",
      primaryClass   = "hep-ph",
      SLACcitation   = "%%CITATION = ARXIV:1411.3739;%%"
}

@article{Nussinov:1985xr,
    author = "Nussinov, S.",
    title = "{Technocosmology: could a technibaryon excess provide a 'natural' missing mass candidate?}",
    reportNumber = "CLNS-85/703",
    doi = "10.1016/0370-2693(85)90689-6",
    journal = "Phys. Lett. B",
    volume = "165",
    pages = "55--58",
    year = "1985"
}

@article{Bagnasco:1993st,
    author = "Bagnasco, John and Dine, Michael and Thomas, Scott D.",
    title = "{Detecting technibaryon dark matter}",
    eprint = "hep-ph/9310290",
    archivePrefix = "arXiv",
    reportNumber = "SCIPP-93-33",
    doi = "10.1016/0370-2693(94)90830-3",
    journal = "Phys. Lett. B",
    volume = "320",
    pages = "99--104",
    year = "1994"
}

@article{Ibe:2018juk,
    author = "Ibe, Masahiro and Kamada, Ayuki and Kobayashi, Shin and Nakano, Wakutaka",
    title = "{Composite Asymmetric Dark Matter with a Dark Photon Portal}",
    eprint = "1805.06876",
    archivePrefix = "arXiv",
    primaryClass = "hep-ph",
    doi = "10.1007/JHEP11(2018)203",
    journal = "JHEP",
    volume = "11",
    pages = "203",
    year = "2018"
}

@article{Wise:2014ola,
      author         = "Wise, Mark B. and Zhang, Yue",
      title          = "{Yukawa Bound States of a Large Number of Fermions}",
      journal        = "JHEP",
      volume         = "02",
      year           = "2015",
      pages          = "023",
      doi            = "10.1007/JHEP10(2015)165, 10.1007/JHEP02(2015)023",
      note           = "[Erratum: JHEP10,165(2015)]",
      eprint         = "1411.1772",
      archivePrefix  = "arXiv",
      primaryClass   = "hep-ph",
      reportNumber   = "CALT-TH-2014-160",
      SLACcitation   = "%%CITATION = ARXIV:1411.1772;%%"
}

@article{Gresham:2017zqi,
      author         = "Gresham, Moira I. and Lou, Hou Keong and Zurek, Kathryn
                        M.",
      title          = "{Nuclear Structure of Bound States of Asymmetric Dark
                        Matter}",
      journal        = "Phys. Rev.",
      volume         = "D96",
      year           = "2017",
      number         = "9",
      pages          = "096012",
      doi            = "10.1103/PhysRevD.96.096012",
      eprint         = "1707.02313",
      archivePrefix  = "arXiv",
      primaryClass   = "hep-ph",
      SLACcitation   = "%%CITATION = ARXIV:1707.02313;%%"
}

@article{Bramante:2024hbr,
    author = "Bramante, Joseph and Diamond, Melissa D. and Kim, J. Leo",
    title = "{Dimming Starlight with Dark Compact Objects}",
    eprint = "2409.08322",
    archivePrefix = "arXiv",
    primaryClass = "hep-ph",
    doi = "10.1103/PhysRevLett.134.141001",
    journal = "Phys. Rev. Lett.",
    volume = "134",
    number = "14",
    pages = "141001",
    year = "2025"
}

@article{Croon:2020wpr,
    author = "Croon, Djuna and McKeen, David and Raj, Nirmal",
    title = "{Gravitational microlensing by dark matter in extended structures}",
    eprint = "2002.08962",
    archivePrefix = "arXiv",
    primaryClass = "astro-ph.CO",
    doi = "10.1103/PhysRevD.101.083013",
    journal = "Phys. Rev. D",
    volume = "101",
    number = "8",
    pages = "083013",
    year = "2020"
}

@article{Bai:2020jfm,
    author = "Bai, Yang and Long, Andrew J. and Lu, Sida",
    title = "{Tests of Dark MACHOs: Lensing, Accretion, and Glow}",
    eprint = "2003.13182",
    archivePrefix = "arXiv",
    primaryClass = "astro-ph.CO",
    doi = "10.1088/1475-7516/2020/09/044",
    journal = "JCAP",
    volume = "09",
    pages = "044",
    year = "2020"
}

@article{Moore:2024mot,
    author = "Moore, Marianne and Slatyer, Tracy R.",
    title = "{Cosmology and terrestrial signals of sexaquark dark matter}",
    eprint = "2403.03972",
    archivePrefix = "arXiv",
    primaryClass = "hep-ph",
    reportNumber = "MIT-CTP/5685",
    doi = "10.1103/PhysRevD.110.023515",
    journal = "Phys. Rev. D",
    volume = "110",
    number = "2",
    pages = "023515",
    year = "2024"
}

@article{Gresham:2017cvl,
      author         = "Gresham, Moira I. and Lou, Hou Keong and Zurek, Kathryn
                        M.",
      title          = "{Early Universe synthesis of asymmetric dark matter
                        nuggets}",
      journal        = "Phys. Rev.",
      volume         = "D97",
      year           = "2018",
      number         = "3",
      pages          = "036003",
      doi            = "10.1103/PhysRevD.97.036003",
      eprint         = "1707.02316",
      archivePrefix  = "arXiv",
      primaryClass   = "hep-ph",
      SLACcitation   = "%%CITATION = ARXIV:1707.02316;%%"
}

@article{Bramante:2017obj,
      author         = "Bramante, Joseph and Unwin, James",
      title          = "{Superheavy Thermal Dark Matter and Primordial
                        Asymmetries}",
      journal        = "JHEP",
      volume         = "02",
      year           = "2017",
      pages          = "119",
      doi            = "10.1007/JHEP02(2017)119",
      eprint         = "1701.05859",
      archivePrefix  = "arXiv",
      primaryClass   = "hep-ph",
      SLACcitation   = "%%CITATION = ARXIV:1701.05859;%%"
}

@article{Affleck:1984fy,
      author         = "Affleck, Ian and Dine, Michael",
      title          = "{A New Mechanism for Baryogenesis}",
      journal        = "Nucl. Phys.",
      volume         = "B249",
      year           = "1985",
      pages          = "361-380",
      doi            = "10.1016/0550-3213(85)90021-5",
      reportNumber   = "Print-84-0574 (PRINCETON)",
      SLACcitation   = "%%CITATION = NUPHA,B249,361;%%"
}

@article{Gresham:2018anj,
      author         = "Gresham, Moira I. and Lou, Hou Keong and Zurek, Kathryn
                        M.",
      title          = "{Astrophysical Signatures of Asymmetric Dark Matter Bound
                        States}",
      journal        = "Phys. Rev.",
      volume         = "D98",
      year           = "2018",
      number         = "9",
      pages          = "096001",
      doi            = "10.1103/PhysRevD.98.096001",
      eprint         = "1805.04512",
      archivePrefix  = "arXiv",
      primaryClass   = "hep-ph",
      SLACcitation   = "%%CITATION = ARXIV:1805.04512;%%"
}

@article{Petraki:2013wwa,
      author         = "Petraki, Kalliopi and Volkas, Raymond R.",
      title          = "{Review of asymmetric dark matter}",
      journal        = "Int. J. Mod. Phys.",
      volume         = "A28",
      year           = "2013",
      pages          = "1330028",
      doi            = "10.1142/S0217751X13300287",
      eprint         = "1305.4939",
      archivePrefix  = "arXiv",
      primaryClass   = "hep-ph",
      reportNumber   = "NIKHEF-2013-016",
      SLACcitation   = "%%CITATION = ARXIV:1305.4939;%%"
}

@article{Zurek:2013wia,
      author         = "Zurek, Kathryn M.",
      title          = "{Asymmetric Dark Matter: Theories, Signatures, and
                        Constraints}",
      journal        = "Phys. Rept.",
      volume         = "537",
      year           = "2014",
      pages          = "91-121",
      doi            = "10.1016/j.physrep.2013.12.001",
      eprint         = "1308.0338",
      archivePrefix  = "arXiv",
      primaryClass   = "hep-ph",
      SLACcitation   = "%%CITATION = ARXIV:1308.0338;%%"
}

@article{Lee:2013bua,
    author = "Lee, Hyun Min and Park, Myeonghun and Sanz, Veronica",
    title = "{Gravity-mediated (or Composite) Dark Matter}",
    eprint = "1306.4107",
    archivePrefix = "arXiv",
    primaryClass = "hep-ph",
    reportNumber = "CERN-PH-TH-2013-143, KIAS-P13032",
    doi = "10.1140/epjc/s10052-014-2715-8",
    journal = "Eur. Phys. J. C",
    volume = "74",
    pages = "2715",
    year = "2014"
}

@article{Hardy:2015boa,
      author         = "Hardy, Edward and Lasenby, Robert and March-Russell, John
                        and West, Stephen M.",
      title          = "{Signatures of Large Composite Dark Matter States}",
      journal        = "JHEP",
      volume         = "07",
      year           = "2015",
      pages          = "133",
      doi            = "10.1007/JHEP07(2015)133",
      eprint         = "1504.05419",
      archivePrefix  = "arXiv",
      primaryClass   = "hep-ph",
      SLACcitation   = "%%CITATION = ARXIV:1504.05419;%%"
}

@article{Kribs:2009fy,
    author = "Kribs, Graham D. and Roy, Tuhin S. and Terning, John and Zurek, Kathryn M.",
    title = "{Quirky Composite Dark Matter}",
    eprint = "0909.2034",
    archivePrefix = "arXiv",
    primaryClass = "hep-ph",
    reportNumber = "FERMILAB-PUB-09-425-T",
    doi = "10.1103/PhysRevD.81.095001",
    journal = "Phys. Rev. D",
    volume = "81",
    pages = "095001",
    year = "2010"
}

@article{Detmold:2014qqa,
      author         = "Detmold, William and McCullough, Matthew and Pochinsky,
                        Andrew",
      title          = "{Dark Nuclei I: Cosmology and Indirect Detection}",
      journal        = "Phys. Rev.",
      volume         = "D90",
      year           = "2014",
      number         = "11",
      pages          = "115013",
      doi            = "10.1103/PhysRevD.90.115013",
      eprint         = "1406.2276",
      archivePrefix  = "arXiv",
      primaryClass   = "hep-ph",
      reportNumber   = "MIT-CTP-4554",
      SLACcitation   = "%%CITATION = ARXIV:1406.2276;%%"
}

@article{Krnjaic:2014xza,
      author         = "Krnjaic, Gordan and Sigurdson, Kris",
      title          = "{Big Bang Darkleosynthesis}",
      journal        = "Phys. Lett.",
      volume         = "B751",
      year           = "2015",
      pages          = "464-468",
      doi            = "10.1016/j.physletb.2015.11.001",
      eprint         = "1406.1171",
      archivePrefix  = "arXiv",
      primaryClass   = "hep-ph",
      SLACcitation   = "%%CITATION = ARXIV:1406.1171;%%"
}

@article{Alves:2009nf,
      author         = "Alves, Daniele S. M. and Behbahani, Siavosh R. and
                        Schuster, Philip and Wacker, Jay G.",
      title          = "{Composite Inelastic Dark Matter}",
      journal        = "Phys. Lett.",
      volume         = "B692",
      year           = "2010",
      pages          = "323-326",
      doi            = "10.1016/j.physletb.2010.08.006",
      eprint         = "0903.3945",
      archivePrefix  = "arXiv",
      primaryClass   = "hep-ph",
      reportNumber   = "SLAC-PUB-14773, SU-ITP-09-13",
      SLACcitation   = "%%CITATION = ARXIV:0903.3945;%%"
}

@article{Wise:2014jva,
    author = "Wise, Mark B. and Zhang, Yue",
    title = "{Stable Bound States of Asymmetric Dark Matter}",
    eprint = "1407.4121",
    archivePrefix = "arXiv",
    primaryClass = "hep-ph",
    reportNumber = "CALT-TH-2014-145",
    doi = "10.1103/PhysRevD.90.055030",
    journal = "Phys. Rev. D",
    volume = "90",
    number = "5",
    pages = "055030",
    year = "2014",
    note = "[Erratum: Phys.Rev.D 91, 039907 (2015)]"
}

@article{Bramante:2019yss,
    author = "Bramante, Joseph and Kumar, Jason and Raj, Nirmal",
    title = "{Dark matter astrometry at underground detectors with multiscatter events}",
    eprint = "1910.05380",
    archivePrefix = "arXiv",
    primaryClass = "hep-ph",
    doi = "10.1103/PhysRevD.100.123016",
    journal = "Phys. Rev. D",
    volume = "100",
    number = "12",
    pages = "123016",
    year = "2019"
}

@article{Bai:2018dxf,
    author = "Bai, Yang and Long, Andrew J. and Lu, Sida",
    title = "{Dark Quark Nuggets}",
    eprint = "1810.04360",
    archivePrefix = "arXiv",
    primaryClass = "hep-ph",
    reportNumber = "FERMILAB-PUB-18-600-T",
    doi = "10.1103/PhysRevD.99.055047",
    journal = "Phys. Rev. D",
    volume = "99",
    number = "5",
    pages = "055047",
    year = "2019"
}

@article{Bai:2019ogh,
    author = "Bai, Yang and Berger, Joshua",
    title = "{Nucleus Capture by Macroscopic Dark Matter}",
    eprint = "1912.02813",
    archivePrefix = "arXiv",
    primaryClass = "hep-ph",
    reportNumber = "PITT-PACC-1912",
    doi = "10.1007/JHEP05(2020)160",
    journal = "JHEP",
    volume = "05",
    pages = "160",
    year = "2020"
}

@article{Hardy:2016kme,
    author = "Hardy, Edward and Lasenby, Robert",
    title = "{Stellar cooling bounds on new light particles: plasma mixing effects}",
    eprint = "1611.05852",
    archivePrefix = "arXiv",
    primaryClass = "hep-ph",
    doi = "10.1007/JHEP02(2017)033",
    journal = "JHEP",
    volume = "02",
    pages = "033",
    year = "2017"
}

@article{Farhi:1984qu,
    author = "Farhi, Edward and Jaffe, R.L.",
    title = "{Strange Matter}",
    reportNumber = "MIT-CTP-1160",
    doi = "10.1103/PhysRevD.30.2379",
    journal = "Phys. Rev. D",
    volume = "30",
    pages = "2379",
    year = "1984"
}

@article{DeRujula:1984axn,
    author = "De Rujula, A. and Glashow, S.L.",
    title = "{Nuclearites: A Novel Form of Cosmic Radiation}",
    reportNumber = "HUTP-84-A057",
    doi = "10.1038/312734a0",
    journal = "Nature",
    volume = "312",
    pages = "734--737",
    year = "1984"
}

@article{Witten:1984rs,
    author = "Witten, Edward",
    title = "{Cosmic Separation of Phases}",
    reportNumber = "PRINT-84-0400 (IAS,PRINCETON)",
    doi = "10.1103/PhysRevD.30.272",
    journal = "Phys. Rev. D",
    volume = "30",
    pages = "272--285",
    year = "1984"
}

@article{Goodman:1984dc,
    author = "Goodman, Mark W. and Witten, Edward",
    editor = "Srednicki, M.A.",
    title = "{Detectability of Certain Dark Matter Candidates}",
    reportNumber = "Print-85-0030 (PRINCETON)",
    doi = "10.1103/PhysRevD.31.3059",
    journal = "Phys. Rev. D",
    volume = "31",
    pages = "3059",
    year = "1985"
}

@article{Drukier:1986tm,
    author = "Drukier, A.K. and Freese, Katherine and Spergel, D.N.",
    title = "{Detecting Cold Dark Matter Candidates}",
    doi = "10.1103/PhysRevD.33.3495",
    journal = "Phys. Rev. D",
    volume = "33",
    pages = "3495--3508",
    year = "1986"
}

@article{Feng:2008mu,
    author = "Feng, Jonathan L. and Tu, Huitzu and Yu, Hai-Bo",
    title = "{Thermal Relics in Hidden Sectors}",
    eprint = "0808.2318",
    archivePrefix = "arXiv",
    primaryClass = "hep-ph",
    reportNumber = "UCI-TR-2008-26",
    doi = "10.1088/1475-7516/2008/10/043",
    journal = "JCAP",
    volume = "10",
    pages = "043",
    year = "2008"
}

@article{Feng:2008ya,
    author = "Feng, Jonathan L. and Kumar, Jason",
    title = "{The WIMPless Miracle: Dark-Matter Particles without Weak-Scale Masses or Weak Interactions}",
    eprint = "0803.4196",
    archivePrefix = "arXiv",
    primaryClass = "hep-ph",
    reportNumber = "UCI-TR-2008-10",
    doi = "10.1103/PhysRevLett.101.231301",
    journal = "Phys. Rev. Lett.",
    volume = "101",
    pages = "231301",
    year = "2008"
}

@article{Pospelov:2007mp,
    author = "Pospelov, Maxim and Ritz, Adam and Voloshin, Mikhail B.",
    title = "{Secluded WIMP Dark Matter}",
    eprint = "0711.4866",
    archivePrefix = "arXiv",
    primaryClass = "hep-ph",
    doi = "10.1016/j.physletb.2008.02.052",
    journal = "Phys. Lett. B",
    volume = "662",
    pages = "53--61",
    year = "2008"
}

@article{Evans:2017kti,
    author = "Evans, Jared A. and Gori, Stefania and Shelton, Jessie",
    title = "{Looking for the WIMP Next Door}",
    eprint = "1712.03974",
    archivePrefix = "arXiv",
    primaryClass = "hep-ph",
    doi = "10.1007/JHEP02(2018)100",
    journal = "JHEP",
    volume = "02",
    pages = "100",
    year = "2018"
}

@article{Acevedo:2024lyr,
    author = "Acevedo, Javier F. and Boukhtouchen, Yilda and Bramante, Joseph and Cappiello, Christopher and Mohlabeng, Gopolang and Tyagi, Narayani",
    title = "{Loosely bound composite dark matter}",
    eprint = "2408.03983",
    archivePrefix = "arXiv",
    primaryClass = "hep-ph",
    doi = "10.1088/1475-7516/2025/03/013",
    journal = "JCAP",
    volume = "03",
    pages = "013",
    year = "2025"
}

@article{Das:2021drz,
    author = "Das, Anirban and Ellis, Sebastian A. R. and Schuster, Philip C. and Zhou, Kevin",
    title = "{Stellar Shocks from Dark Matter Asteroid Impacts}",
    eprint = "2106.09033",
    archivePrefix = "arXiv",
    primaryClass = "hep-ph",
    reportNumber = "SLAC-PUB-17604",
    doi = "10.1103/PhysRevLett.128.021101",
    journal = "Phys. Rev. Lett.",
    volume = "128",
    number = "2",
    pages = "021101",
    year = "2022"
}

@article{Heeba:2018wtf,
    author = {Heeba, Saniya and Kahlhoefer, Felix and St\"ocker, Patrick},
    title = "{Freeze-in production of decaying dark matter in five steps}",
    eprint = "1809.04849",
    archivePrefix = "arXiv",
    primaryClass = "hep-ph",
    reportNumber = "TTK-18-38",
    doi = "10.1088/1475-7516/2018/11/048",
    journal = "JCAP",
    volume = "11",
    pages = "048",
    year = "2018"
}

@article{Bramante:2016yju,
    author = "Bramante, Joseph and Cook, Jessica and Delgado, Antonio and Martin, Adam",
    title = "{Low Scale Inflation at High Energy Colliders and Meson Factories}",
    eprint = "1608.08625",
    archivePrefix = "arXiv",
    primaryClass = "hep-ph",
    doi = "10.1103/PhysRevD.94.115012",
    journal = "Phys. Rev. D",
    volume = "94",
    number = "11",
    pages = "115012",
    year = "2016"
}

@article{Husdal:2016haj,
    author = "Husdal, Lars",
    title = "{On Effective Degrees of Freedom in the Early Universe}",
    eprint = "1609.04979",
    archivePrefix = "arXiv",
    primaryClass = "astro-ph.CO",
    doi = "10.3390/galaxies4040078",
    journal = "Galaxies",
    volume = "4",
    number = "4",
    pages = "78",
    year = "2016"
}

@article{Planck:2018vyg,
    author = "Aghanim, N. and others",
    collaboration = "Planck",
    title = "{Planck 2018 results. VI. Cosmological parameters}",
    eprint = "1807.06209",
    archivePrefix = "arXiv",
    primaryClass = "astro-ph.CO",
    doi = "10.1051/0004-6361/201833910",
    journal = "Astron. Astrophys.",
    volume = "641",
    pages = "A6",
    year = "2020",
    note = "[Erratum: Astron.Astrophys. 652, C4 (2021)]"
}

@article{CMB-S4:2016ple,
    author = "Abazajian, Kevork N. and others",
    collaboration = "CMB-S4",
    title = "{CMB-S4 Science Book, First Edition}",
    eprint = "1610.02743",
    archivePrefix = "arXiv",
    primaryClass = "astro-ph.CO",
    reportNumber = "FERMILAB-FN-1024-A-AE",
    month = "10",
    year = "2016"
}

@article{Lasenby,
    author = "Hardy, Edward and Lasenby, Robert",
    title = "{Stellar cooling bounds on new light particles: plasma mixing effects}",
    eprint = "1611.05852",
    archivePrefix = "arXiv",
    primaryClass = "hep-ph",
    doi = "10.1007/JHEP02(2017)033",
    journal = "JHEP",
    volume = "02",
    pages = "033",
    year = "2017"
}

@article{Ellis:2008hf,
    author = "Ellis, John R. and Olive, Keith A. and Savage, Christopher",
    title = "{Hadronic Uncertainties in the Elastic Scattering of Supersymmetric Dark Matter}",
    eprint = "0801.3656",
    archivePrefix = "arXiv",
    primaryClass = "hep-ph",
    reportNumber = "CERN-PH-TH-2008-005, UMN-TH-2631-08, FTPI-MINN-08-02",
    doi = "10.1103/PhysRevD.77.065026",
    journal = "Phys. Rev. D",
    volume = "77",
    pages = "065026",
    year = "2008"
}

@article{Lebedev:2021xey,
    author = "Lebedev, Oleg",
    title = "{The Higgs portal to cosmology}",
    eprint = "2104.03342",
    archivePrefix = "arXiv",
    primaryClass = "hep-ph",
    doi = "10.1016/j.ppnp.2021.103881",
    journal = "Prog. Part. Nucl. Phys.",
    volume = "120",
    pages = "103881",
    year = "2021"
}

@article{Silveira:1985rk,
    author = "Silveira, Vanda and Zee, A.",
    title = "{SCALAR PHANTOMS}",
    reportNumber = "DOE-ER-40048-13 P5",
    doi = "10.1016/0370-2693(85)90624-0",
    journal = "Phys. Lett. B",
    volume = "161",
    pages = "136--140",
    year = "1985"
}

@article{ff_constraint_2,
    author = "Decca, R. S. and Lopez, D. and Fischbach, E. and Klimchitskaya, G. L. and Krause, D. E. and Mostepanenko, V. M.",
    title = "{Novel constraints on light elementary particles and extra-dimensional physics from the Casimir effect}",
    eprint = "0706.3283",
    archivePrefix = "arXiv",
    primaryClass = "hep-ph",
    doi = "10.1140/epjc/s10052-007-0346-z",
    journal = "Eur. Phys. J. C",
    volume = "51",
    pages = "963--975",
    year = "2007"
}

@article{ff_constraint_4,
  title = {Test of the Gravitational Inverse Square Law at Millimeter Ranges},
  author = {Yang, Shan-Qing and Zhan, Bi-Fu and Wang, Qing-Lan and Shao, Cheng-Gang and Tu, Liang-Cheng and Tan, Wen-Hai and Luo, Jun},
  journal = {Phys. Rev. Lett.},
  volume = {108},
  issue = {8},
  pages = {081101},
  numpages = {5},
  year = {2012},
  month = {Feb},
  publisher = {American Physical Society},
  doi = {10.1103/PhysRevLett.108.081101},
  url = {https://link.aps.org/doi/10.1103/PhysRevLett.108.081101}
}

@article{ff_constraint_3,
    author = "Sushkov, A. O. and Kim, W. J. and Dalvit, D. A. R. and Lamoreaux, S. K.",
    title = "{New Experimental Limits on Non-Newtonian Forces in the Micrometer Range}",
    eprint = "1108.2547",
    archivePrefix = "arXiv",
    primaryClass = "quant-ph",
    doi = "10.1103/PhysRevLett.107.171101",
    journal = "Phys. Rev. Lett.",
    volume = "107",
    pages = "171101",
    year = "2011"
}

@article{ff_constraint_1,
    author = "Adelberger, E. G. and Heckel, Blayne R. and Nelson, A. E.",
    title = "{Tests of the gravitational inverse square law}",
    eprint = "hep-ph/0307284",
    archivePrefix = "arXiv",
    doi = "10.1146/annurev.nucl.53.041002.110503",
    journal = "Ann. Rev. Nucl. Part. Sci.",
    volume = "53",
    pages = "77--121",
    year = "2003"
}

@article{Xie:2024mxr,
    author = "Xie, Ke-Pan",
    title = "{Revisiting the fermion-field nontopological solitons}",
    eprint = "2405.01227",
    archivePrefix = "arXiv",
    primaryClass = "hep-ph",
    doi = "10.1007/JHEP09(2024)077",
    journal = "JHEP",
    volume = "09",
    pages = "077",
    year = "2024"
}

@article{Bottaro:2023gep,
    author = "Bottaro, Salvatore and Caputo, Andrea and Raffelt, Georg and Vitagliano, Edoardo",
    title = "{Stellar limits on scalars from electron-nucleus bremsstrahlung}",
    eprint = "2303.00778",
    archivePrefix = "arXiv",
    primaryClass = "hep-ph",
    reportNumber = "CERN-TH-2023-035",
    doi = "10.1088/1475-7516/2023/07/071",
    journal = "JCAP",
    volume = "07",
    pages = "071",
    year = "2023"
}

@article{Goudzovski:2022vbt,
    author = "Goudzovski, Evgueni and others",
    title = "{New physics searches at kaon and hyperon factories}",
    eprint = "2201.07805",
    archivePrefix = "arXiv",
    primaryClass = "hep-ph",
    reportNumber = "FERMILAB-PUB-22-057-T",
    doi = "10.1088/1361-6633/ac9cee",
    journal = "Rept. Prog. Phys.",
    volume = "86",
    number = "1",
    pages = "016201",
    year = "2023"
}

@article{Flacke:2016szy,
    author = "Flacke, Thomas and Frugiuele, Claudia and Fuchs, Elina and Gupta, Rick S. and Perez, Gilad",
    title = "{Phenomenology of relaxion-Higgs mixing}",
    eprint = "1610.02025",
    archivePrefix = "arXiv",
    primaryClass = "hep-ph",
    reportNumber = "CTPU-16-25",
    doi = "10.1007/JHEP06(2017)050",
    journal = "JHEP",
    volume = "06",
    pages = "050",
    year = "2017"
}

@article{Fradette:2018hhl,
    author = "Fradette, Anthony and Pospelov, Maxim and Pradler, Josef and Ritz, Adam",
    title = "{Cosmological beam dump: constraints on dark scalars mixed with the Higgs boson}",
    eprint = "1812.07585",
    archivePrefix = "arXiv",
    primaryClass = "hep-ph",
    doi = "10.1103/PhysRevD.99.075004",
    journal = "Phys. Rev. D",
    volume = "99",
    number = "7",
    pages = "075004",
    year = "2019"
}
\end{document}